\documentclass[aip, amsmath, amssymb, reprint,floatfix]{revtex4-2}
\usepackage[utf8]{inputenc}
\usepackage{graphicx} 

\usepackage[hidelinks]{hyperref}
\usepackage{xcolor}
\date{September 2021}
\newcommand{\br}{\mathbf{r}}
\newcommand{\bv}{\mathbf{v}}

\newcommand{\bE}{\mathbf{E}}
\newcommand{\bB}{\mathbf{B}}

\newcommand{\bj}{\mathbf{j}}
\newcommand{\bR}{\mathbf{R}}

\newcommand{\tr}{\tilde{r}}

\newcommand{\tp}{\tilde{p}}
\newcommand{\tx}{\tilde{x}}
\newcommand{\tn}{\tilde{n}}
\newcommand{\tT}{\tilde{T}}

\newcommand{\tv}{\tilde{v}}

\newcommand{\tB}{\tilde{B}}

\newcommand{\tE}{\tilde{E}}

\newcommand{\tm}{\tilde{m}}
\newcommand{\pdv}[2]{\frac{\partial #1}{\partial #2}}

\newcommand{\tomega}{\tilde{\omega}}

\begin{document}
\preprint{AIP/123-QED}

\title{Finite-Difference Multiple Fluid Solution for Source-Driven Rotation in Highly Magnetized Linear Plasma Device}
\author{T. Rubin}
\email{trubin@princeton.edu}
\affiliation{Department of Astrophysical Sciences, Princeton University, Princeton, New Jersey 08544, USA}

\author{E. J. Kolmes}
\affiliation{Department of Astrophysical Sciences, Princeton University, Princeton, New Jersey 08544, USA}
\author{I. E. Ochs}
\affiliation{Department of Astrophysical Sciences, Princeton University, Princeton, New Jersey 08544, USA}
\author{M. E. Mlodik}
\affiliation{Department of Astrophysical Sciences, Princeton University, Princeton, New Jersey 08544, USA}
\author{N. J. Fisch}
\affiliation{Department of Astrophysical Sciences, Princeton University, Princeton, New Jersey 08544, USA}

\begin{abstract}
    The rotation profile of a magnetized plasma cylinder composed of multiple fluids is investigated analytically, expanding on previous results. The analytic steady-state solution is used as a benchmark for a time-dependent multiple-fluid finite-difference code, MITNS: Multiple-Ion Transport Numerical Solver. Magnetic field evolution is taken into account, both analytically and numerically. Its details are shown to be of importance when particles are allowed out of the domain. MITNS reproduces the asymptotic expansion results for a small parameter $\delta\lll1$. For $\sqrt{m_e/m_i} \sim \delta \ll 1$, a slightly different regime, dominated by viscosity-induced transport of ions, is found numerically and analytically. This validation supports the use of this code for more complex time-dependent calculations in the future.
\end{abstract}
\maketitle

\noindent
\maketitle
\section{Introduction}~\label{sec:1}
A cylindrical plasma geometry is used in fusion applications such as mirrors~\cite{Post1987,Gueroult2012}, z-pinches~\cite{Velikovich}, MagLIF~\cite{Knapp2019,Gomez2019,Slutz2012,Gomez2014}, and in some centrifugal fusion concepts~\cite{bekhtenev1980problems,Ellis2001,Fetterman2008,Kolmes2018,Rax2017}, as well as mass separation applications such as plasma centrifuges~\cite{Bonnevier1966,Lehnert1971,Hellsten1977,Krishnan1983,Geva1984,Dolgolenko2017,Ochs2017iii,Yuferov2018,Gueroult2018,Fetterman2011b,Ohkawa2002,ONeil1981,Rax2016,Gueroult2018ii}. Investigation of classical transport effects is a first step in modeling of these devices.

Literature regarding classical transport in plasma, employing Braginskii~\cite{Braginskii1965} transport equations, or equivalent formulations for multi-species multi-fluid plasmas~\cite{Zhdanov_2002} is extensive, containing predictions for flow profiles~\cite{Litvinova2021}, currents~\cite{Radial_Current}, and impurity pinches~\cite{Strategies,Mlodik2021} expected in multi-fluid plasmas. The rotation profile in such devices affects the centrifugal force and shear stress, and through them the ion density profiles~\cite{Strategies}, and viscous heating~\cite{HotIonMode}. Radial and azimuthal currents depend on the rotation profile as well~\cite{Radial_Current}. Differential rotation of ion species results in ion-ion frictional heating and enhanced heat transport~\cite{Mlodik2020} (Ettingshausen effect).

A multiple-fluid model of classical cross-field transport involves solving $4N$ partial differential equations, for the densities, momenta and energies for the $N$ fluids. This equation set is complemented by $14N$ boundary conditions ($2N$ particle fluxes, $4N$ momenta fluxes, $2N$ energy fluxes, in addition to $4N$ momenta diffusion terms, and $2N$ heat diffusion terms), in addition to $4N$ possible volumetric source terms, such as particle injection, wave-driven body forces or laser heating. The evolution of such plasma is nonlinear and complex.

Tailoring the rotation profile might result in enhanced device performance, but the number of boundary conditions and source terms requires a numerical tool able to take these terms into account all together.

There are many plasma simulation codes. Some solve the Braginskii single ion and electron fluid~\cite{ricci2012}, or add neutral species~\cite{leake2014,laguna2016}. Some solve anomalous transport \textit{N}-fluid models with a common temperature profile~\cite{rognlien19942,simonini1994,radford1996,braams1996,wiesen2015,rognlien2018}, and others solve \textit{N}-fluid unmagnetized plasma~\cite{rambo1994,rambo1995,ghosh2019}.

MITNS: Multiple-Ion Transport Numerical Solver~\cite{MITNS}, is just such a numerical tool, solving an arbitrary number of one-dimensional coupled ion-fluid equations - for species specific density, velocity and pressure - in addition to Faraday's law for magnetic field evolution, and the electron pressure equation. Electron density and velocity are taken into account using an MHD-like approximation, which enforces quasi-neutrality to leading order, and drops the inertial terms. 

In this paper, we extend the slab code MITNS to a cylindrical geometry, with spatial derivatives allowed only in the radial coordinate. This code is used to numerically validate previous results~\cite{Radial_Current}, discuss the validity of some assumptions present there, and explore a different equation ordering.

This code is capable of evolving self-consistently not only the continuity and momentum equations, but also the pressure equation for all species and the magnetic field equation. This enables detailed simulations, accounting for the differences in actuation mechanisms (e.g. boundary driven systems vs. source driven systems, etc.), and the different resulting regimes (e.g. magnetic field dynamics, heating). 

Potential use cases for this code are the time evolution of laser-heated plasma columns, which partition the heat to the electrons more so than the ions, and the tracking of impurities in such scenarios. Or solving for the heat dissipation and pressure buildup of counter-flowing fluxes, as would occur in steady-state fusion devices. 

This paper is organized as follows: in Sec~\ref{sec:2}, we present the model equations. In Sec~\ref{sec:3} we discuss the previous solution, and add to it the magnetic field equation. In Sec~\ref{sec:4} we compare simulation results to a steady-state solution, and in Sec~\ref{sec:5} we re-derive the rotation profile for a multi-species plasma containing several types of ions, in either a cylindrical or annular domain, for arbitrary boundary conditions.




\section{Multiple Fluid Equations for Cylindrical Plasma Device}~\label{sec:2}

In this section, we present the model equations for an \textit{N}-fluid system with imposed temperature profiles, using the closure by \citet{Zhdanov_2002}.  The treatment is confined to cylindrical coordinates, with gradients only in the radial direction. Using these equations, a leading order solution to the angular velocity profile, density and magnetic field is derived.

For a fluid plasma composed of several species, the continuity and momentum equations for each species are,
\begin{gather}
    \pdv{n_s}{t} + \nabla \cdot (n_s \bv_s) = s_s,~\label{eq:continuity}
\end{gather}
\begin{multline}
    \pdv{}{t}(m_s n_s \bv_s) + \nabla \cdot\left(m_s n_s \bv_s \bv_s \right) +\nabla \cdot \pi_s + \nabla p_s\\=  Z_s e n_s \left( \bE + \bv_s \times \bB \right)  +m_s s_s \bv_s^{src}\\+  \sum_{s'}\left(\bR_{ss'} +\mathbf{f}_{ss'}\right),~\label{eq:momentum}
\end{multline}
with the friction force $\bR_{ss'}$ and the thermal friction (``Nernst") force $\mathbf{f}_{ss'}$ between species $s$ and $s'$,
\begin{gather}
    \bR_{ss'} = m_s n_s \nu_{ss'}(\bv_{s'}-\bv_{s})\\
    \mathbf{f}_{ss'} = \frac{3}{2}\frac{m_s n_s \nu_{ss'}}{Z_s Z_{s'} e B}\hat b\times\frac{Z_{s'}m_{s'}T_s\nabla T_s - Z_{s}m_{s}T_{s'}\nabla T_{s'} }{m_s T_{s'}+m_{s'}T_s}\label{eq:Nernst}.
\end{gather}

Here, we shall consider gradients only in the $\hat r$ direction for an axisymmetric cylindrical (or annular) geometry, $\pdv{}{\theta} = \pdv{}{z} = 0$. The magnetic field is taken to be in the perpendicular, $\hat z$ direction. The resultant viscous stress tensor divergence, using the \citet{Braginskii1965} and \citet{Zhdanov_2002} closure is,

\begin{multline}
    \nabla \cdot \pi_s = \pdv{}{r}\left[\frac{\eta_{s0}}{3r}\pdv{}{r}\left(rv_{sr}\right)\right]\hat r\\-\frac{1}{r^2}\frac{\partial }{\partial r}r^3\left[\eta_{s1}  \frac{\partial }{\partial r}\left( \frac{v_{sr}}{r} \right)+\eta_{s3} \frac{\partial }{\partial r}\left( \frac{v_{s\theta}}{r} \right) \right] \hat r\\ -\frac{1}{r^2}\frac{\partial }{\partial r}  r^3 \left[\eta_{s1}  \frac{\partial }{\partial r}\left( \frac{v_{s\theta}}{r} \right) -\eta_{s3}  \frac{\partial }{\partial r}\left( \frac{v_{sr}}{r} \right) \right]\hat \theta\label{eq:divpi}.
\end{multline}
This viscous stress tensor corresponds to the multi-species generalization to Braginskii's~\cite{Braginskii1965} closure by Zhdanov~\cite{Zhdanov_2002}, with coefficients,
\begin{gather}
    \eta_{s1} = \frac{p_s}{4\Omega_s^2}\sum_{s'}\frac{\sqrt{2}m_sm_{s'}\nu_{ss'}}{(m_s+m_{s'})^2}\left(\frac{6}{5}\frac{m_{s'}}{m_s}+2-\frac{4}{5}\frac{\Omega_s}{\Omega_{s'}}\right),\\
    \eta_{s3} = \frac{p_s}{2\Omega_s}.
\end{gather}

In Cartesian coordinates, the divergence of the viscous stress is a diffusion of linear momentum, $\mathbf{P}_s = m_s n_s \bv_s$. The second and third lines in equation (\ref{eq:divpi}) take the form of a diffusion term for $\br \times \mathbf{P}_s$, the angular momentum, and $\br \cdot \mathbf{P}_s$, the mass flux. Specifically, it is not a diffusion term for $m_s n_s v_{sr}$ and $m_s n_s v_{s\theta}$. More details are discussed in section \ref{sec:5}. 

We leave the pressure equation out in this instance, and determine the pressure $p_s = n_s T_s$ from the density, and an imposed temperature profile $T_s = T_s(r,t)$.

The magnetic field evolution is determined by Faraday's Law,
\begin{gather}
    \pdv{\bB}{t} = -\nabla \times \bE~\label{eq:Faraday}.
\end{gather}

Alternatively, the magnetic field can be determined in steady state from Ampere's law,
\begin{gather}
    \nabla \times \bB = \mu_0 \bj.
\end{gather}

\section{Source-Driven Rotation}~\label{sec:3}

A recent paper,~\citet{Radial_Current} ordered the steady-state velocity terms in equations (\ref{eq:momentum}) in powers of
\begin{gather}
    \delta \doteq \frac{E}{r\Omega_i B},\label{eq:delta}\\
    \epsilon \doteq \frac{\nu_{ie}}{\Omega_i},
\end{gather}
with respect to $\bE\times \bB$, the leading order flow velocity, assumed by Braginskii. Their solution to the (single) ion and electron fluids was expressed in terms of the ion particle flux,
\begin{gather}
    \Gamma_i(r) \doteq r n_i v_{ir} = \Gamma_i(r_i)+\int_{r_i}^{r} r s_i dr,
\end{gather}
and the electric charge source,
\begin{gather}
    C(r) \doteq r j_r = e\sum_s Z_s \Gamma_s.
\end{gather}

One might note that in cylindrical geometry with $\pdv{}{t} = \pdv{}{z} = \pdv{}{\theta} = 0$, electric charge conservation dictates $C \equiv 0$. The case  $C\neq 0 $ can only be interpreted as a proxy for a charge transport process outside of the radial classical transport scope of this work, such as a kinetic or wave-driven phenomenon. Alternatively, one can think of it as a proxy for the axial variation, see appendix \ref{Appendix:proxy}.

The ODEs describing the leading order dimensionless rotation $\tilde \omega_{rot}$ and density profiles have been derived, (equations (28) and (48) in~\cite{Radial_Current}): 

\begin{gather}
    \tilde \omega_{rot}\doteq -\frac{E}{rB}\frac{R}{v_{thi}}\\
    \tr^3\tilde \eta \tilde\omega_{rot}' =\tr^2\mathcal{QP}\tilde \Gamma\tilde \omega_{rot}+ \mathcal{I}\int_0^{\tr} \tr' \tilde C  \tB(\tr') d\tr' \label{eq:omegaprime},\\
    \mathcal{P}\tilde \Gamma =\tr\frac{\tn_i\tn_e }{\tT_e^{3/2}\tB^2 } \left[\left(1+\frac{Z_i m_e}{m_i}\right)\tr\tilde \omega_{rot}^2 -\frac{\tp_i'+\tp_e'}{  \tn_i} +\frac{3}{2}Z_i \tT_e'  \right] \label{eq:Si},
\end{gather}
with equation (\ref{eq:Si}) being a correction to equation (28) in~\cite{Radial_Current}, in which the $\tr \tomega_{rot}^2$ term is erroneously multiplied by $\tT_i$.

The dimensionless viscosity, viscosity ratio, particle and source and current, as well as the dimensionless primary small parameter, are given by
\begin{gather}
    \tilde \eta \doteq \frac{\tn}{\sqrt{\tT_i}\tB^2},\\
    \mathcal{Q} \doteq \frac{10\sqrt{2}\nu_{ie0}}{3\nu_{ii0}}, 
    \\
    \mathcal{P} \doteq  \frac{1}{n_0 \rho_{i0}^2 \nu_{ie0}}\Gamma(R),~\label{eq:n_0}
    \\
    \mathcal{I} \doteq \frac{10\sqrt{2} R }{3 Z_i e n_0 \rho_{i0}^3  \nu_{ii0}}C(R),
    \\
    \delta = \frac{\rho_{i0}}{R}\frac{\tE}{\tr \tB^2} = \tilde \rho_{i0}\frac{\tE}{\tr \tB^2}.
\end{gather}

The solution to equation (\ref{eq:omegaprime}) is given as,
\begin{multline}
    \tilde\omega_{rot}(\tr) = e^{\mathcal{QP}\int_1^{\tr}(\Tilde \Gamma / \tx \eta) d\tx} \Bigg[\tilde\omega_{rot}\big|_{\tr=1}\\+\mathcal{I}\int_1^{\tr}\frac{e^{-\mathcal{QP}\int_1^{\tr'}(\Tilde \Gamma / \tx \eta) d\tx}}{\tr'^3 \eta(\tr')}\int_0^{\tr'} \tilde y \tilde C(\tilde y) \tB(\tilde y)d\tilde y \tr'\Bigg]. \label{eq:omega_sol}
\end{multline}

\begin{figure*}[t]
    \centering
    \includegraphics[width=\textwidth]{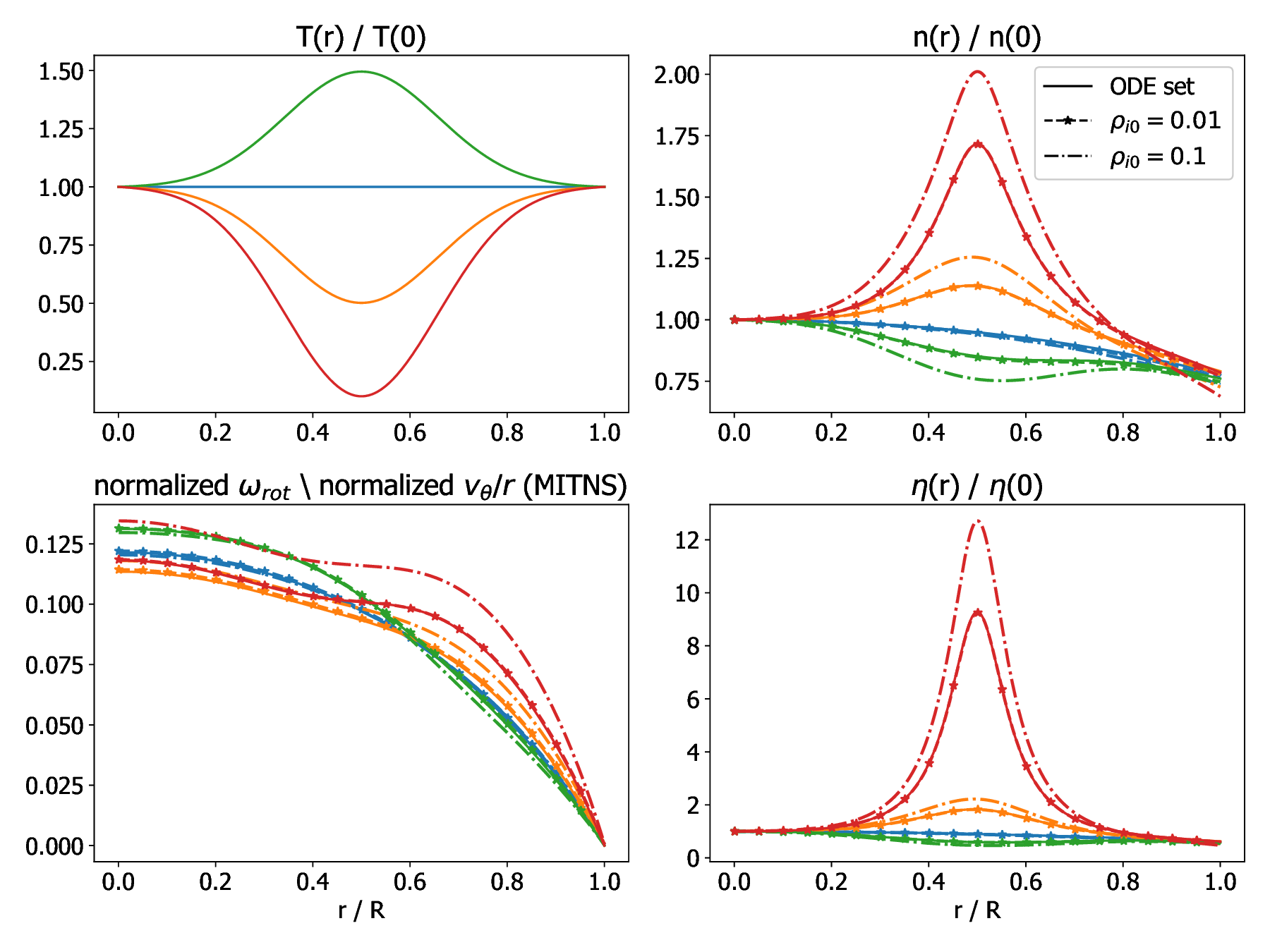}
    \caption{Several solutions of the rotation profile, density and viscosity coefficient corresponding to some steady state temperature profiles, $\mathcal{P} = 1$, $\mathcal{I} = -1$, $\mathcal{Q} = 0.1556$. Comparison between MITNS results, for reference normalized ion Larmor radius $\tilde \rho_{i0} = 0.01,\ 0.1$. Markers for the $\tilde \rho_{i0} = 0.01$ are placed every $30$ grid points for visibility. Notice the agreement between the normalized (first order) angular velocity from the ODE set (full line) and the full rotation frequency from MITNS (dashed line with markers).}
    \label{fig:MITNSComp}
\end{figure*}

One might notice that in order to keep $\mathcal P, \mathcal I \sim \mathcal O (1)$, the source terms scale as $\tilde s_i \sim \tilde \rho_{p0}^2$, $\sum_s Z_s\tilde s_s \sim \tilde \rho_{p0}^3$.

Using Ampere's law, we can derive the ODE for the magnetic field strength to the same order of accuracy,
\begin{gather}
    \tB' = \frac{\tn_i \tr \tilde \omega_{rot}^2-\left(\tp_i'+\tp_e'\right)}{\tv_A^2\tB},~\label{eq:B_0}
\end{gather}
with the normalized Alfv\'en speed,
\begin{gather}
    \tv_A^2 \doteq \frac{B_0^2}{n_0 T_0 \mu_0}.
\end{gather}

Equation (\ref{eq:B_0}) can be integrated to yield,
\begin{gather}
    \tB(\tr) = \sqrt{\tB^2\big|_{\tr=0}+\frac{2}{\tv_A^2}\int_0^{\tr} \left(\tn_i \tr \tilde \omega_{rot}^2-\left(\tp_i'+\tp_e'\right)\right)d\tr}~\label{eq:Bsol}.
\end{gather}
where the constant of integration might be selected to satisfy $\int_0^1\tB \tr d\tr = 1/2$, corresponding to a radial rearrangement of a magnetic field with initial uniform strength $\tB = 1$.

The boundary condition for (\ref{eq:n_0}) might be also selected for a set particle number $\tilde N = \int_0^1\tn_i \tr d\tr = 1/2$, also corresponding to a radial rearrangement of a uniform particle density $\tn_i = 1$.

\begin{figure}[!t]
   \centering
    \includegraphics[width=\columnwidth]{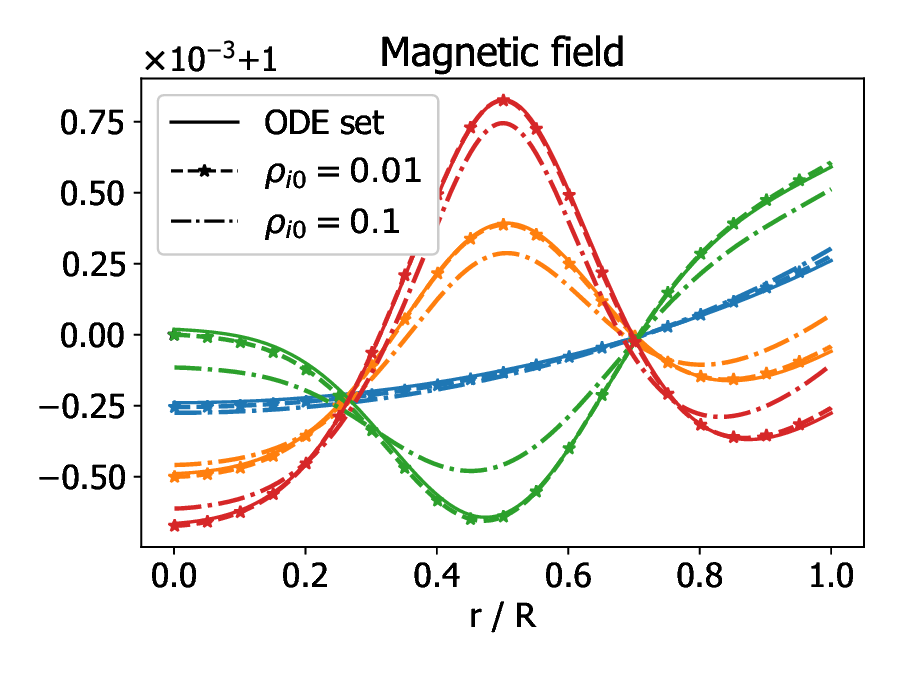}
    \caption{Comparison between equation (\ref{eq:Bsol}) and MITNS solution for the dimensionless magnetic field, using $\tv_A^2 = 1000$.}
    \label{fig:MITNSB}
\end{figure}

\citet{Radial_Current} plot (Figure 3) the steady state normalized density, angular velocity and viscosity distributions for four temperature profiles, for the case of uniform ion an electron particle sources ($\tilde \Gamma = \tr^2, \tilde C = \tr^2$), with $\mathcal{P} = 1$, $\mathcal{I}=-1$ and $\mathcal{Q} = 0.1556$ - corresponding to $H^+$ ions. That figure contains an error which Figure \ref{fig:MITNSComp} corrects.

\section{Full simulation in MITNS}~\label{sec:4}

As mentioned above, MITNS~\cite{MITNS} is a numerical tool, capable of evolving the first three fluid moments for an arbitrary number of ion species, as well as Faraday's law. In order to test the predictions in~\citet{Radial_Current}, a cylindrical coordinate mode was added to MITNS. In the simulations described in this paper, the pressure equation was disabled, and a temperature was prescribed, $T_i = T_e = T(r,t)$.

MITNS takes electrons into account only algebraically, as discussed in appendix~\ref{Appendix:RadCurr}, and so the electron inertial terms in the momentum equation do not appear. In terms of agreement to the model presented here, the $\tr \tilde \omega_{rot}^2$ term in equation (\ref{eq:Si}), would not have the $Z_i m_e/m_i$ term upfront. Similarly, the electron viscosity has been neglected, as it is smaller by a factor of $m_e/m_i$ in relation to the ion viscosity.

MITNS evaluates $n_s$, $v_{s\theta}$, $p_s$, and $B_z$ and their time derivatives at cell-centers, and as such avoids the coordinate singularity at $r=0$, in the divergence terms $\mathrm{div}\ \mathbf{V} = \frac{1}{r}\pdv{}{r} rV_r$, and $(\mathrm{div}\ \mathbf{T})\cdot\hat \theta =\frac{1}{r}\left[\pdv{}{r} rT_{r\theta}+T_{\theta r}\right]$. For $v_r$, since $v_r(0,t) = 0$ due to axisymmetry, there is no need to evaluate $\pdv{}{t}v_r(0,t)$. For the viscous stress tensor, writing
\begin{gather}
    r^3\pdv{}{r}\frac{v_\theta}{r} = r^2 v_\theta' - rv_\theta 
\end{gather}
eliminates the need to evaluate $v_\theta/r$ at $r=0$ in order to compute the viscous stress on the cell adjacent to $r=0$.

A monotonized-central Van-Leer~\cite{VANLEER1979101} flux limiter is employed for the particle and momentum fluxes, as well as the magnetic field flux, in an upwind scheme, in order to avoid spurious oscillations.

In order to establish density steady state, the ion flux at the outer radius was set to be equal to the integrated ion source term, leading to no change in total particle number within the domain.

The magnetic field evolution MITNS is determined by Faraday's law,
\begin{multline}
    \pdv{B_z}{t} =-\frac{1}{r}\pdv{}{r}rE_\theta=\\ -\frac{1}{r}\pdv{}{r}r\left[v_{er}B_z + \sum_s\left(\frac{m_e \nu_{es}}{e} (v_{s\theta} - v_{e\theta})+\frac{f_{es}}{en_e}\right)\right],\label{eq:dbdt}
\end{multline}
where the electric field is determined by the electron momentum equation, as seen in the second line of equation (\ref{eq:dbdt}).

In steady state, the $v_{er}B_z$ term and the friction and Nernst force terms balance each other. In this simulation, particles are injected in to the simulation domain, which generates radial particle fluxes. These radial particle fluxes generate torques, which then produce angular velocity in the fluids. This means that before steady state is established, the advection term, which depends on the radial electron flux, is larger than the azimuthal term in (\ref{eq:dbdt}), and magnetic field is pushed out initially. 

Without imposing zero magnetic field flux out of the outer radius, this would lead to some magnetic field leaving the computation domain, and thus being lost. Electron-ion friction provides magnetic field diffusion, and prevents it from piling up at edge of the domain in steady state. 


It is possible to have an annular domain, and, for example, inject particles only through the inner boundary instead of volumetrically. In that case, an appropriate set of coils around the inner electrode could ensure the edge source of electrons carry magnetic field into the domain, which would replenish the initial magnetic field loss. This setup would produce a different torque on the plasma $\tau = \br\times(\bj\times\bB)\propto B_z$ rather than $\propto r^2B_z$, which would be balanced by a different viscous stress profile, and hence a different angular velocity profile.

For the results presented here, a zero advective magnetic field flux boundary condition was set on the outer radius of the domain, such that the total magnetic flux in the domain remains constant.
\subsection{Comparison of Analytic Transport Theory with MITNS}

A comparison between the solution to equations (\ref{eq:omegaprime}-\ref{eq:Si}), produced as an asymptotic expansion in small parameters $\delta,\ \epsilon$, and the steady-state results of a simulation in MITNS is presented in Figure \ref{fig:MITNSComp}. The small parameter $\tilde \rho_{i0}$ is set by specifying the domain size in MITNS, and $\delta$ equation (\ref{eq:delta}),  depends on it, together with the electric and magnetic fields produced as a result of the applied sources and boundary conditions. 
A comparison between the magnetic field given by equation (\ref{eq:Bsol}), and the one solved for in MITNS is shown in Figure \ref{fig:MITNSB}. 

\begin{figure}[t]
    \centering
    \includegraphics[width=\columnwidth]{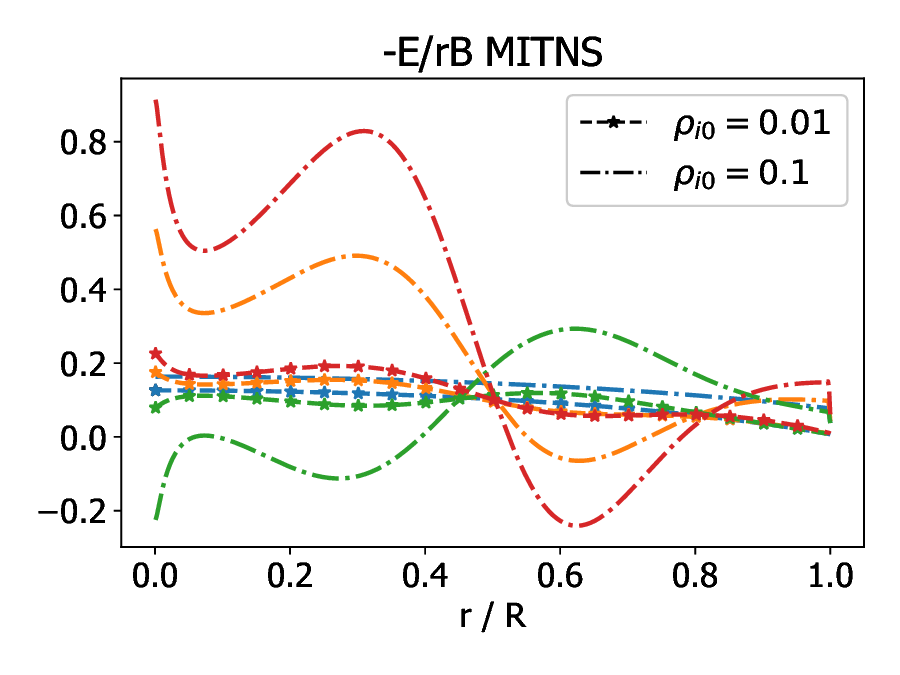}
    \caption{MITNS calculated dimensionless $-E/rB$ for the same cases presented in Figure \ref{fig:MITNSComp}. Large disagreements between the dimensionless first order rotation frequency in figure \ref{fig:MITNSComp}, and the resultant electric field here.}
    \label{fig:MITNSErB}
\end{figure}

There is a good agreement between the ODE set solution and the steady-state MITNS result for $\tilde \rho_{i0} = 0.01$,as evidenced by the closeness of the dashed and full lines in Figures \ref{fig:MITNSComp} and \ref{fig:MITNSB}. The small parameter, $|\delta(r)|$ in this case was $<0.002$. For  $\tilde \rho_{i0} = 0.1$, $|\delta|\lessapprox0.06\sim \sqrt{m_e / m_i}$, the agreement is not as good.  

\begin{figure}[t]
    \centering
    \includegraphics[width=\columnwidth]{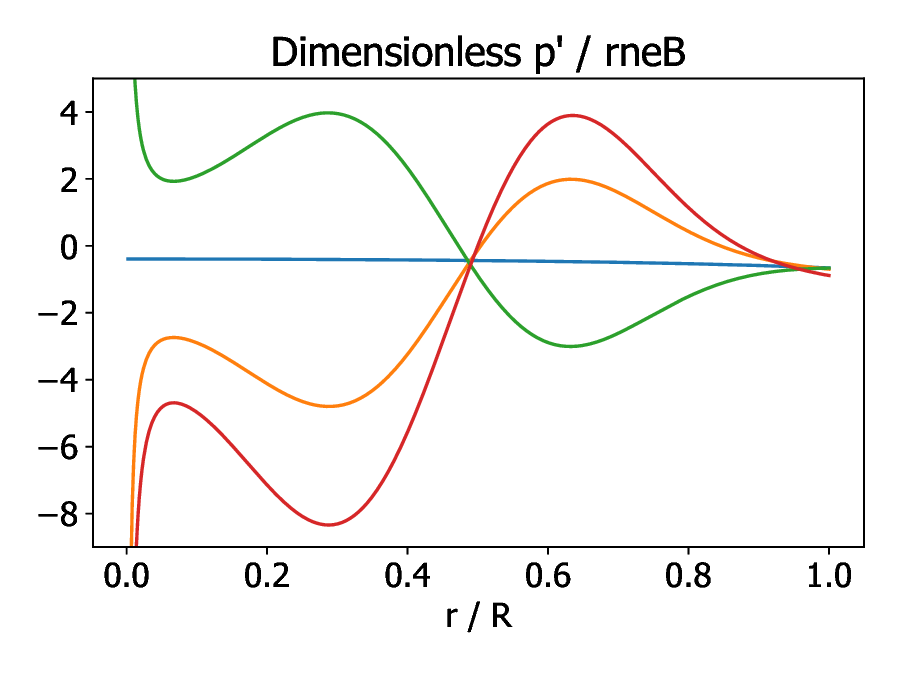}
    \caption{ODE set calculated dimensionless diamagnetic angular velocity. 
    }
    \label{fig:diamag}
\end{figure}

\section{Angular-Momentum Transport View of Radial Currents}~\label{sec:5}

Interestingly, comparing the expression $-E/rB$ evaluated by MITNS, as shown in figure \ref{fig:MITNSErB}, to the angular velocity profile in figure \ref{fig:MITNSComp} shows there are some differences between them, when solving the system of equations for a small but finite $\tilde \rho_{i0}$. The $\bE\times\bB$ azimuthal drift discussed in \citet{Radial_Current} is recovered in the limit of $\tilde \rho_{i0}\rightarrow0$. 

For the $\tilde \rho_{i0}=0.01$ case, the most visible features in figure \ref{fig:MITNSErB}, are the non-zero derivative at $r=0$, and the non-monotonicity of the green curve, even though the angular velocity profile is monotonous in figure \ref{fig:MITNSComp}. This means that the solution to equation (\ref{eq:omegaprime}) contains information about the diamagnetic drift as well as $-E/rB$.

For the $\tilde \rho_{i0}=0.1$ case, the diamagnetic drift $p'/neB$, which was assumed to be $\sim\mathcal O (\delta E/B)$, becomes leading-order, as the pressure varies across a much smaller domain. The diamagnetic angular frequency,
\begin{gather}
    \frac{p'}{rneB} = \frac{T_0}{R^2 e B_0}\frac{\tp'}{\tr\tn\tB} = \frac{v_{thi}}{R}\tilde \rho_{i0}\frac{\tp'}{\tr\tn\tB},
\end{gather}
with the dimensionless $\tp'/\tr \tn\tB$ (plotted in Figure \ref{fig:diamag}), is $\sim \mathcal O (3)$. In order to make it an order smaller than $\tomega$ (lower-left panel in Figure \ref{fig:MITNSComp}), a $\tilde \rho_{i0}\ll 0.12/3\sim 0.04$ is required.

In this section, we aim to expand upon the ordering of the equations of motion for \textit{N}-fluids, derive the angular velocity and density profiles in terms of the cross-field particle fluxes, and show the solution presented in \ref{sec:3} is a particular solution obtained for a specific ordering of the particle flux magnitude. The small parameters in the expansion presented in this section are formally constants, rather than a function of the solution, which is easier to handle. Additionally, we explicitly keep the inner boundary terms such as in equation (\ref{eq:vr}), (\ref{eq:totalrot}) and (\ref{eq:totradial}), which are set to zero in \citet{Radial_Current} equation (48), for example. These boundary terms would be useful in annular geometries used for homopolar generators or some mass filter applications.

Writing the equations of motion in term of angular momentum, rather than linear momentum, casts the viscous torque term in a diffusion form. This form enables the angular momentum equation to be written in a finite-volume form, which is computationally conservative scheme.

The continuity equation, and the radial particle flux, 
\begin{gather}
    \pdv{n_s}{t} + \frac{1}{r}\pdv{}{r}\Gamma_s = s_s,\\
    \Gamma_s = r n_s v_{sr} = \Gamma_s(r_i)+\int_{r_i}^{r}\left(s_s - \pdv{n_s}{t}\right)r'dr'~\label{eq:vr},
\end{gather}
are the driving mechanisms for the system dynamics, and they are a reasonable boundary condition to experimentally impose.

The angular momentum conservation equation,
\begin{gather}
    \ell_{s} \doteq r m_s n_s v_{s\theta} = \br \times \mathbf{P}_{s}\cdot \hat z,
\end{gather}
\begin{multline} 
    \pdv{\ell_{s}}{t} + \frac{1}{r}\pdv{}{r} \frac{\ell_{s} \Gamma_s}{n_s} +Z_s e \Gamma_s B_z =  \sum_{s'}\left(\nu_{s's}\ell_{s'}-\nu_{ss'}\ell_s\right)\\
    +\frac{1}{r}\pdv{}{r}r^3 \left[\eta_{s1}\omega_s'-\eta_{s3}\pdv{}{r}\left(\frac{\Gamma_s}{r^2n_s}\right)\right] +r\sum_{s'}f_{ss'\theta}+\ell_s^{src},~\label{eq:l}
\end{multline}
is written here as an advection-diffusion equation for each species, with source terms corresponding to the Lorentz torque on the left hand side, and the friction and Nernst torques on the right hand side. The last term, $\ell^{src}_s$ indicates other sources of angular momentum, such as the injection of particles with non-zero angular velocity. In that case, $\ell^{src}_s = r^2 m_s s_s \omega_{s}^{src}$. We shall take this term to be zero from now.

The radial component of the momentum equation, for $P_{sr} \doteq m_s n_s v_{sr}$,
\begin{multline} 
    \pdv{P_{sr}}{t} + \frac{1}{r}\pdv{}{r}rP_{sr}v_{sr} +p_s'= \frac{1}{r}\ell_s\omega_s+Z_sen_s\left(E_r+r\omega_s B_z\right)\\+\sum_{s'}\left(\nu_{s's}P_{s'r}-\nu_{ss'}P_{sr}\right)\\+\frac{1}{r^2}\frac{\partial }{\partial r}r^3\left[\eta_{s1}  \frac{\partial }{\partial r}\left( \frac{v_{sr}}{r} \right)+\eta_{s3}\omega_s' \right]+P_{sr}^{src},
\end{multline}
can be written in a form such that the viscosity is a diffusion term. This allows for a conservative-form solution, as already discussed. 

Using
\begin{gather}
    M_{s}\doteq r m_s n_s v_{sr} = \br \cdot \mathbf{P}_s = m_s \Gamma_s,
\end{gather}
the conservation equation for $M_{s}$ is
\begin{multline} 
    \pdv{M_{s}}{t} + \frac{1}{r}\pdv{}{r}\left(\frac{M_{s}\Gamma_{s}}{n_s} + r^2 p_s\right)= 2p_s + \frac{M_{s}\Gamma_{s}}{r^2n_s} + \omega_s\ell_{s}\\+Z_sen_sr\left(E_r+r\omega_s B_z\right)+\sum_{s'}\left(\nu_{s's}M_{s'}-\nu_{ss'}M_{s}\right)\\+\frac{1}{r}\frac{\partial }{\partial r}r^3\left[\eta_{s1}  \frac{\partial }{\partial r}\left( \frac{v_{sr}}{r} \right)+\eta_{s3}\omega_s' \right]+M_{s}^{src}.\label{eq:M}
\end{multline}

In steady state, the integral from of the two equations, (\ref{eq:M}), (\ref{eq:l}):
\begin{multline} 
    \frac{\ell_{s} \Gamma_s}{n_s}\Big|_{r_i}^{r} +Z_s e \int_{r_i}^{r}\Gamma_s B_zr'dr' = \\ \sum_{s'}\int_{r_i}^{r}\left(\nu_{s's}\ell_{s'}-\nu_{ss'}\ell_s+f_{ss'\theta}r'\right)r'dr'\\
    +r^3 \left[\eta_{s1}\omega_s'-\eta_{s3}\pdv{}{r}\left(\frac{\Gamma_s}{r^2n_s}\right)\right]_{r_i}^{r},
\end{multline}
\begin{multline} 
    \left[\frac{M_{s}\Gamma_{s}}{n_s} + r^2 p_s\right]_{r_i}^{r}= \int_{r_i}^{r}\left(2p_s + \frac{M_{s}\Gamma_{s}}{r'^2n_s} + \omega_s\ell_{s}\right)r'dr'\\+Z_se\int_{r_i}^{r} n_s\left(E_r+r'\omega_s B_z\right)r'^2dr'\\+\sum_{s'}\int_{r_i}^{r}\left(\nu_{s's}M_{s'}-\nu_{ss'}M_{s}\right)r'dr'\\+r^3\left[\eta_{s1}  \frac{\partial }{\partial r}\left( \frac{\Gamma_s}{r^2 n_s} \right)+\eta_{s3}\omega_s' \right]_{r_i}^{r},
\end{multline}
where integrals of the form $\int_{r_i}^{r}\cdot \ r' dr'$ should de understood as volume average in the annular volume bounded by $r_i$ and $r$.

Summing over all species, the friction forces cancel, and the electric field term drops off due to quasi-neutrality.
\begin{multline} 
    \sum_s m_s r^2 \omega_s \Gamma_s\Big|_{r_i}^{r} + \int_{r_i}^{r}C B_zr'dr' = \\\sum_sr^3 \left[\eta_{s1}\omega_s'-\eta_{s3}\pdv{}{r}\left(\frac{\Gamma_s}{r^2n_s}\right)\right]_{r_i}^{r},\label{eq:totalrot}
\end{multline}
\begin{multline} 
    \sum_s\left[m_s\frac{\Gamma_{s}^2}{n_s} + r^2 p_s\right]_{r_i}^{r}=\sum_sZ_se\int_{r_i}^{r} n_s\omega_s B_z r'^3dr'\\+ \sum_s\int_{r_i}^{r}\left(2p_s + m_s\frac{\Gamma_{s}^2}{r'^2n_s} + \omega_s\ell_{s}\right)r'dr'\\+\sum_s r^3\left[\eta_{s1}  \frac{\partial }{\partial r}\left( \frac{\Gamma_s}{r^2 n_s} \right)+\eta_{s3}\omega_s' \right]_{r_i}^{r}.\label{eq:totradial}
\end{multline}

These equations do not depend on the electric field, and we will show their equivalence to equations (\ref{eq:omegaprime}) and (\ref{eq:Si}). The leading order $\omega$, even though it must be of $\mathcal O \left(\frac{E}{rB}\right)$, by the choice of the Braginskii transport coefficients, may contain information from other particle drifts.

\subsection{Nondimensionalization}
Nondimensionalizing the equations of motion would factor out small parameters that would be used in an asymptotic expansion.

Denoting $X = X_0 \tilde X$, with $X_0$ being a reference quantity; 
\begin{gather}
    m_0 = m_p,\\
    v_0 = \sqrt{T_0/m_p},\\
    r_0 = R,\\
    \Gamma_0 = R n_0 v_0,\\
    C_0 = e R n_0 v_0,\\
    \eta_{30} = \frac{n_0T_0}{\Omega_{p0}},\\
    \eta_{10} = \frac{n_0T_0\nu_0}{\Omega_{p0}^2},\\
    R_{ss'0}= m_0 n_0 \nu_0 v_{0},\\
    f_{ss'0} = \frac{n_0 \nu_0 T_0}{\Omega_{p0} R} = m_0 n_0 \nu_0 v_0 \tilde \rho_{p0}\label{eq:NernstDimen}
\end{gather}
with $R$ being the outer domain radius.

The dimensionless equations feature the two small parameters $\tilde \rho_{p0} = v_0/ \Omega_{p0}R$, and  $\epsilon = \nu_0 / \Omega_{p0}$. They are ordered such that $1\gg \epsilon \gg \tilde \rho_{p0}$.

The steady-state dimensionless angular momentum for a single fluid species is,
\begin{multline}
     \tm_s \tr^2 \tilde\Gamma_s \tomega_s\Big|_{\tr_i}^{\tr} +\frac{1}{\tilde \rho_{p0}}\int_{\tr_i}^{\tr}Z_s \tr' \tilde\Gamma_s \tB_zd\tr'=\\\tilde \rho_{p0}\tr^3 \left[\epsilon\tilde \eta_{s1}\tomega_s'-\tilde\eta_{s3}\pdv{}{\tr}\left(\frac{\tilde \Gamma_s}{\tr^2 \tn_s}\right)\right]\Bigg|_{\tr_i}^{\tr}
    \\+\epsilon \int_{\tr_i}^{\tr}\sum_{s'}\left(\frac{1}{\tilde \rho_{p0}}\tilde R_{ss'\theta} +\tilde f_{ss'\theta}\right)\tr'^2d\tr'.\label{eq:dimSingleRot}
\end{multline}

The equation for the entire plasma is,
\begin{multline}
    \sum_s\tr^2\tm_s\Gamma_s\tomega_s \Big|_{\tr_i}^{\tr} = \tilde \rho_{p0}\epsilon \tr^3 \sum_s  \tilde \eta_{s1}\tomega_s'\Big|_{\tr_i}^{\tr}\\-\tilde \rho_{p0} \tr^3\sum_s \tilde \eta_{s3}\pdv{}{\tr}\left(\frac{\tilde \Gamma_s}{\tr^2 \tn_s}\right)\Bigg|_{\tr_i}^{\tr}
    -\frac{1}{\tilde \rho_{p0}}\int_{\tr_i}^{\tr} \tr' \tilde C \tB_z d\tr'.\label{eq:dimexactrot}
\end{multline}

We shall order the angular velocity and source terms in the two small parameters using $\tilde X = \sum_{\alpha, \beta}\tilde \rho_{p0}^{\alpha}\epsilon^\beta \tilde X^{(\alpha,\beta)}$, and restrict $\tomega_s$ such that its leading term is $\tomega_s^{(0,0)}$.


Equation (\ref{eq:dimSingleRot}), when expanded to $\mathcal O (\tilde \rho_{p0}^k)$, $k=-1,0$ yield,
\begin{gather}
    \int_{\tr_i}^{\tr}Z_s \tr' \tilde\Gamma_s^{(k+1,0)} \tB_zd\tr'=0,\\
    \tilde\Gamma_s^{(0,0)} =\tilde\Gamma_s^{(1,0)}=0.
\end{gather}

To $\mathcal{O}(\tilde \rho_{p0}^{-1}\epsilon)$,
\begin{gather}
    Z_s\tilde\Gamma_s^{(0,1)} \tB_z= \sum_{s'}\left( \tm_s \tn_s\tilde\nu_{ss'}\tr(\tomega_{s'}^{(0,0)}-\tomega_{s}^{(0,0)})  +\tilde f_{ss'\theta}^{(-1,0)}\right)\tr.
\end{gather}
Here, $\tilde f_{ss'\theta}^{(-1,0)}$ is understood as the dimensionless (\ref{eq:Nernst}) with dimensionless temperature gradients of $\mathcal{O}(\tilde \rho_{p0}^{-1})$.

To $\mathcal{O}(\epsilon)$:
\begin{multline}
    \int_{\tr_i}^{\tr}\sum_{s'}\left(\tr\tm_s \tn_s \tilde \nu_{ss'}(\tomega_{s'}^{(1,0)}-\tomega_s^{(1,0)}) +\tilde f_{ss'\theta}^{(0,0)}\right)\tr'^2d\tr'
    \\=\tr^2 \tm_s \left[\tilde\Gamma_s^{(0,1)}\tomega_s^{(0,0)}\right]_{\tr_i}^{\tr} +\int_{r_i}^{r}Z_s \tr' \tilde\Gamma_s^{(1,1)} \tB_zd\tr'.\label{eq:SIIIII}
\end{multline}
This equation is the equivalent to equation (\ref{eq:Si}), assuming the leading order $\tilde \Gamma_s$ is $\tilde \Gamma_s^{(1,1)}$.


Equation (\ref{eq:dimexactrot}), when expanded to $\mathcal O (\tilde \rho_{p0}^k)$, $k=-1,0,1$ yield,
\begin{gather}
    \int_{\tr_i}^{\tr} \tr' \tilde C^{(k+1,0)} \tB_z d\tr'=0,\\
    \tilde C^{(0,0)} =\tilde C^{(1,0)} = \tilde C^{(2,0)}=0.
\end{gather}

and when expanded to $\mathcal O (\tilde \rho_{p0}\epsilon)$,
\begin{multline}
    \tr^2\sum_s\tm_s\left[\tilde\Gamma_s^{(1,1)}\tomega_s^{(0,0)}+ \tilde\Gamma_s^{(0,1)}\tomega_s^{(1,0)}\right]_{\tr_i}^{\tr}=\\ \tr^3 \sum_s  \tilde \eta_{s1}\tomega_s^{(0,0)'}\Big|_{\tr_i}^{\tr}-\tr^3\sum_s \tilde \eta_{s3}\pdv{}{\tr}\left(\frac{\tilde \Gamma_s^{(0,1)}}{\tr^2 \tn_s}\right)\Bigg|_{\tr_i}^{\tr}\\
    -\int_{\tr_i}^{\tr} \tr' \tilde C^{(2,1)} \tB_z d\tr'.~\label{eq:omega(1,1)}
\end{multline}

Equation (\ref{eq:omega(1,1)}) is the differential equation for $\tomega_s^{(0,0)}$, assuming $\tilde C = \tilde \rho_{p0}^2 \epsilon\tilde C^{(2,1)}$, and $\tilde \Gamma_s = \tilde \rho_{p0} \epsilon \tilde \Gamma_s^{(1,1)}$, are known functions. If $\tr_i = 0$, it reduces to (\ref{eq:omegaprime}).


To $\mathcal O (\tilde \rho_{p0}^2 \epsilon)$,
\begin{multline}
    \tr^2\sum_s \tm_s\left[\tilde \Gamma_s^{(2,1)}\tomega_s^{(0,0)}+\tilde \Gamma_s^{(0,1)}\tomega_s^{(2,0)}+\tilde \Gamma_s^{(1,1)}\tomega_s^{(1,0)}\right]_{\tr_i}^{\tr}\\
    + \tr^3\sum_s \tilde \eta_{s3}\pdv{}{\tr}\left(\frac{\tilde \Gamma_s^{(1,1)}}{\tr^2 \tn_s}\right)\Bigg|_{\tr_i}^{\tr}
    +\int_{\tr_i}^{\tr} \tr' \tilde C^{(3,1)} \tB_z d\tr'= 
    \\ \tr^3 \sum_s\tilde \eta_{s1}\tomega_s^{(1,0)'}  \Big|_{\tr_i}^{\tr}
    -\tr^2\sum_s \tm_s\left[\tilde \Gamma_s^{(2,0)}\tomega_s^{(0,1)}\right]_{\tr_i}^{\tr}.
\end{multline}

This becomes the differential equation for $\tomega_s^{(1,0)}$, assuming $\tilde C = \tilde \rho_{p0}^2 \epsilon\tilde C^{(2,1)}$, and $\tilde \Gamma_s = \tilde \rho_{p0} \epsilon \tilde \Gamma_s^{(1,1)}$, are known functions,
\begin{multline}
    \sum_s \tm_s\left[\tilde \Gamma_s^{(1,1)}\tomega_s^{(1,0)}\right]_{\tr_i}^{\tr}
    + \tr\sum_s \tilde \eta_{s3}\pdv{}{\tr}\left(\frac{\tilde \Gamma_s^{(1,1)}}{\tr^2 \tn_s}\right)\Bigg|_{\tr_i}^{\tr}= 
    \\ \tr \sum_s\tilde \eta_{s1}\tomega_s^{(1,0)'}\Big|_{\tr_i}^{\tr}.
\end{multline}

The dimensionless steady-state mass flux equation for a single fluid species is,
\begin{multline} 
    \left[\tm_s\frac{\tilde \Gamma_{s}^2}{\tn_s} + \tr^2 \tp_s\right]_{\tr_i}^{\tr}= \frac{1}{\tilde \rho_{p0}}Z_s\int_{\tr_i}^{\tr} \tn_s\left[\tE_r+\tr'\tomega_s \tB_z\right]\tr'^2d\tr'\\+\int_{\tr_i}^{\tr}\left(2\tp_s + \tm_s\frac{\tilde \Gamma_{s}^2}{\tr'^2\tn_s} + \tm_s\tn_s\tr'^2\tomega_s^2\right)\tr'd\tr'\\+\frac{\epsilon}{\tilde \rho_{p0}}\sum_{s'}\int_{\tr_i}^{\tr}\left(\tm_{s'}\tilde\nu_{s's}\tilde \Gamma_{s'}-\tm_s \tilde\nu_{ss'}\tilde \Gamma_{s}\right)r'dr'\\+\tilde \rho_{p0}\tr^3\left[\epsilon\tilde\eta_{s1}  \frac{\partial }{\partial \tr}\left( \frac{\tilde\Gamma_s}{\tr^2 \tn_s} \right)+\tilde \eta_{s3}\tomega_s' \right]_{\tr_i}^{\tr},\label{eq:vrssdim}
\end{multline}

To leading order, $\mathcal{O}(\tilde \rho_{p0}^{-1})$,
\begin{gather} 
    \tp_s^{(-1,0)'}= Z_s\tn_s\left(\tE_r^{(0,0)}+\tr\tomega_s^{(0,0)} \tB_z\right),\label{eq:electric}
\end{gather}
where $\tp_s^{(-1,0)'}$ is understood as resulting from temperature gradients that are of $\mathcal{O}(\tilde \rho_{p0}^{-1})$. This is the case in the numerical example in section \ref{sec:4}, when $\tilde \rho_{p0} = 0.1$.

Equation (\ref{eq:electric}) is deceptive, as it seems to imply the leading order rotation is an $\bE\times\bB$ drift. However, one arrives at this equation after solving for $\tomega_s^{(0,0)}$. This equation actually defines the electric field profile. This assertion is reflected in figures \ref{fig:MITNSErB} and \ref{fig:diamag}, in which the sum of the non-monotonous pressure gradient and electric field terms cancel each other such that the remainder is the monotonous angular velocity. In these figures this is the case for both $\tilde \rho_{p0} = 0.1$ and $0.01$.


To $\mathcal{O}(1)$, subtracting equation (\ref{eq:vrssdim}) for $s=s'$ from itself with $s=s$,
\begin{multline}
    \tomega_{s'}^{(1,0)}-\tomega_s^{(1,0)}=\frac{\tp_{s'}'^{(0,0)}}{Z_{s'}\tn_{s'}\tr\tB_z}- \frac{\tm_{s'}\tomega_{s'}^{(0,0)^2}}{Z_{s'}\tB_z}\\-\frac{\tp_s'^{(0,0)}}{Z_s\tn_s\tr\tB_z}+ \frac{\tm_s\tomega_s^{(0,0)^2}}{Z_s\tB_z},
\end{multline}
allows us to achieve a non-ambiguous expression, to be substituted in equation (\ref{eq:SIIIII}).

The dimensionless equation (\ref{eq:totradial}), \begin{multline} 
    \sum_s\left[\tm_s\frac{\tilde\Gamma_{s}^2}{\tn_s} + \tr^2 \tp_s\right]_{\tr_i}^{\tr}=\frac{1}{\tilde \rho_{p0}}\sum_sZ_s\int_{\tr_i}^{\tr} \tn_s\tomega_s \tB_z \tr'^3d\tr'\\+ \sum_s\int_{\tr_i}^{\tr}\left(2\tp_s + \tm_s\frac{\tilde \Gamma_{s}^2}{\tr'^2\tn_s} +\tm_s\tn_s \tomega_s^2\tr'^2\right)\tr'd\tr'\\+\tilde \rho_{p0}\sum_s \tr^3\left[\epsilon\tilde\eta_{s1}  \pdv{}{\tr}\left( \frac{\tilde\Gamma_s}{\tr^2 \tn_s} \right)+\tilde\eta_{s3}\tomega_s' \right]_{\tr_i}^{\tr}.
\end{multline}

To leading order, using Ampere's law, this is an equation for the magnetic field,

\begin{gather}
    \sum_s \tp_s'+  \tr\tomega_s^{(0,0)^2}\sum_s \tm_s \tn_s  =  \mu_0^{-1}\pdv{\tB_z^{(1,0)}}{\tr}\tB_z.
\end{gather} 

In this section we derived the leading and first order correction equations for the angular velocity profile, and the leading order equations for the density and magnetic field. We have shown how different particle flux magnitudes affect the solution, and suggested an interpretation of the relation between the leading order rotation and the electric field.

\section{Conclusion}
The code MITNS: Multiple-Ion Transport Numerical Solver, was expanded to include a cylindrical coordinate mode, and used to validate the first-order solution to the rotation frequency and density distribution in a source-driven, axially magnetized, rotating two-fluid plasma cylinder presented in~\citet{Radial_Current}. 

First, simulation results pointed to the error in the plotting script for Figure 3 in~\cite{Radial_Current}, and we were able to correct it in Figure \ref{fig:MITNSComp}. 

Second, we have shown the approach to steady-state should be performed carefully, with appropriate magnetic boundary conditions, or else particle fluxes out of the simulation domain will deplete the magnetic field. 

Third, after requiring a constant magnetic flux in the cylinder, the results of the MITNS code and the ODE set solution were shown to be congruent in the limit of small Larmor radius over domain size. This lends credibility to the MITNS code and its cylindrical coordinates mode.

Finally, we derived the rotation frequency, density and magnetic field equations, and suggested an interpretation to the relation between the electric field and rotation profile. 

\subsection*{Acknowledgment}
This work was supported by Cornell NNSA 83228-10966 [Prime No. DOE (NNSA) DE-NA0003764] and by NSF-PHY-1805316.
\section*{Author Declarations}
\subsection*{Conflict of nterest}
The authors have no conflicts to disclose.
\subsection*{Data Availability}
The data that support the findings of this study are available from the corresponding author upon reasonable request.
\appendix
\section{Using the particle source-term as a proxy for the axial dimension}~\label{Appendix:proxy}

In a steady state, azimuthally-symmetric cylinder and in the absence of particle source terms, the continuity equation can be written as, 
\begin{gather}
    \nabla \cdot (n_s \bv_s) = \frac{1}{r}\pdv{}{r}(r n_s v_{sr}) + \pdv{}{z}(n_s v_{sz}) = s_s^{\mathrm{source}},
\end{gather}
in the presence of fusion or ionization / recombination processes. $s_s^{\mathrm{source}}$ should be understood as a charge preserving in the sense $\sum_s Z_s s_s^{\mathrm{source}}=0$.

We are interested in the result of radial variation, so we can treat the $\pdv{}{z}$ part of the divergence as an additional source term,
\begin{gather}
    \frac{1}{r}\pdv{}{r}(r n_s v_{sr})  = s_s = s_s^{\mathrm{source}}- \pdv{}{z}(n_s v_{sz}). \label{eq:source_as_pdvz}
\end{gather}
Identifying part of the source term with out-of-plane flow allows for charge / current source, where the electric charge weighted sum of electron and ion sources and sinks does not sum to zero.

Similarly for the momentum equation, we have a degree of freedom in $\bv_s^{src}$. Noticing the $\pdv{}{z}$ component of $n_s \left( \bv_s \cdot \nabla \right) \bv_s$ is $n_s v_z \pdv{\bv_s}{z}$, we can choose $\bv_s^{src}$ such that:
\begin{gather}
    \bv_s^{src} = \bv + \frac{n_s v_{sz}}{\pdv{}{z}(n_s v_{sz})-s_s^{\mathrm{source}}}\pdv{\bv}{z} = \bv - \frac{n_s v_{sz}}{s_s}\pdv{\bv}{z}\label{eq:vel_source_as_pdvz}
\end{gather}

Interpretation of the particle and velocity source terms as equations (\ref{eq:source_as_pdvz}) and (\ref{eq:vel_source_as_pdvz}) allows us to solve for 2d "slices" of a long cylindrical device. This is a method of getting an electric current source term that is large and entirely described by classical transport.

It is also possible to interpret the volumetric radial current source as being produced by a wave-particle interaction. In this picture, the wave is not modeled using the electromagnetic fields, due to its fast time-scales, and only pushes a current. Its ponderomotive force might appear as a momentum source term.

Another interpretation to the current source might be an externally-imposed shift in the magnetic field, which carries with it the electron fluid before friction is able to diffuse the field back.

\section{Resolving radial current in a 1D Magneto-hydrodynamic multiple fluid simulation}~\label{Appendix:RadCurr}

The equations solved in MITNS, (\ref{eq:continuity}) and (\ref{eq:momentum}), are "MHD-like". Fast time-scales are discarded, such as the electric field term in Ampere's law, in addition to the electron density and inertia terms.

Without the time derivatives in the electron continuity and momentum equations, the electron fluid density and velocity is evaluated algebraically from quasi-neutrality,
\begin{gather}
    n_e = \sum_{s\neq e } Z_s n_s,
\end{gather}
and form Ampere's law,
\begin{gather}
    \mu_0 e n_e \bv_e=\mu_0\sum_{s\neq e} Z_s e n_s \bv_s - \nabla \times \bB.
\end{gather}

Quasi-neutrality makes Gauss Law inapplicable for a non-zero electric field. Instead, the electric field is determined from the inertia-less and non-viscous electron momentum equations. This electric field is also used to evolve the magnetic field. 

MITNS has a one-dimensional domain, which restricts the allowed variations to the radial direction. If $\pdv{}{\theta} = \pdv{}{z} =0$,
\begin{gather}
    \nabla \times \bB \cdot \hat r\equiv 0,~\label{eq:nojr}
\end{gather}
and no current can flow in the radial direction. 

In order to explore radial plasma conductivity in this framework, that is, to look at the relation between $j_r$ and the other physical quantities in the system, we must ignore equation (\ref{eq:nojr}), and instead implement a current on the entire domain, similarly to a body force (such as gravity),
\begin{gather}
    v_{er}=\left[\sum_{s\neq e } Z_s n_s\right]^{-1}\cdot\left[\sum_{s\neq e} Z_s  n_s v_{sr} - j_r(r,t)\right].
\end{gather}

The function $j_r(r,t)$ has to be specified either as a local Ohm's law, or as an a-priori driving force.

The algebraic nature of the "MHD-like" electric field prevents an application of boundary conditions to the electron momentum equation or to the electric field. 

Azimuthal current can be injected into the simulated domain by applying a boundary condition to the magnetic field.

This solution adds to the electron velocity perpendicularly to the magnetic field. This does not represents a cross-field transport of the electrons, as the magnetic field would be advected by the moving electron fluid. Adding the current to the electrons is numerically simple, and its effects are automatically implemented in the friction terms and magnetic field evolution. It also avoids changing the accurate ion momentum equations. Attempting to add the imposed current to the ion velocities would require some partition of the current among an arbitrary number of ion fluids.

It is possible to evaluate, in post processing, the departure from quasi-neutrality using Gauss law by taking the divergence of the electric field calculated using Ohm's law. The current relates to the displacement current by, 
\begin{gather}
    \nabla\cdot\bj = -\varepsilon_0 \pdv{}{t}\nabla\cdot \bE,\\
    j_r =\frac{r_i j_r(r_i)}{r} -\varepsilon_0 \pdv{}{t}\left(E_r-\frac{r_i E_r(r_i)}{r}\right),
\end{gather}
which would be a small, of $\mathcal{O}(v_A^2/c^2)$, with $v_A$ being the Alfv\'en speed, and $c$ the speed of light.

\section*{References}

\begin{thebibliography}{48}%
\makeatletter
\providecommand \@ifxundefined [1]{%
 \@ifx{#1\undefined}
}%
\providecommand \@ifnum [1]{%
 \ifnum #1\expandafter \@firstoftwo
 \else \expandafter \@secondoftwo
 \fi
}%
\providecommand \@ifx [1]{%
 \ifx #1\expandafter \@firstoftwo
 \else \expandafter \@secondoftwo
 \fi
}%
\providecommand \natexlab [1]{#1}%
\providecommand \enquote  [1]{``#1''}%
\providecommand \bibnamefont  [1]{#1}%
\providecommand \bibfnamefont [1]{#1}%
\providecommand \citenamefont [1]{#1}%
\providecommand \href@noop [0]{\@secondoftwo}%
\providecommand \href [0]{\begingroup \@sanitize@url \@href}%
\providecommand \@href[1]{\@@startlink{#1}\@@href}%
\providecommand \@@href[1]{\endgroup#1\@@endlink}%
\providecommand \@sanitize@url [0]{\catcode `\\12\catcode `\$12\catcode
  `\&12\catcode `\#12\catcode `\^12\catcode `\_12\catcode `\%12\relax}%
\providecommand \@@startlink[1]{}%
\providecommand \@@endlink[0]{}%
\providecommand \url  [0]{\begingroup\@sanitize@url \@url }%
\providecommand \@url [1]{\endgroup\@href {#1}{\urlprefix }}%
\providecommand \urlprefix  [0]{URL }%
\providecommand \Eprint [0]{\href }%
\providecommand \doibase [0]{https://doi.org/}%
\providecommand \selectlanguage [0]{\@gobble}%
\providecommand \bibinfo  [0]{\@secondoftwo}%
\providecommand \bibfield  [0]{\@secondoftwo}%
\providecommand \translation [1]{[#1]}%
\providecommand \BibitemOpen [0]{}%
\providecommand \bibitemStop [0]{}%
\providecommand \bibitemNoStop [0]{.\EOS\space}%
\providecommand \EOS [0]{\spacefactor3000\relax}%
\providecommand \BibitemShut  [1]{\csname bibitem#1\endcsname}%
\let\auto@bib@innerbib\@empty
\bibitem [{\citenamefont {Post}(1987)}]{Post1987}%
  \BibitemOpen
  \bibfield  {author} {\bibinfo {author} {\bibfnamefont {R.~F.}\ \bibnamefont
  {Post}},\ }\href {https://doi.org/10.1088/0029-5515/27/10/001} {\bibfield
  {journal} {\bibinfo  {journal} {Nucl. Fusion}\ }\textbf {\bibinfo {volume}
  {27}},\ \bibinfo {pages} {1579} (\bibinfo {year} {1987})}\BibitemShut
  {NoStop}%
\bibitem [{\citenamefont {Gueroult}\ and\ \citenamefont
  {Fisch}(2012)}]{Gueroult2012}%
  \BibitemOpen
  \bibfield  {author} {\bibinfo {author} {\bibfnamefont {R.}~\bibnamefont
  {Gueroult}}\ and\ \bibinfo {author} {\bibfnamefont {N.~J.}\ \bibnamefont
  {Fisch}},\ }\href {https://doi.org/10.1063/1.4765692} {\bibfield  {journal}
  {\bibinfo  {journal} {Phys. Plasmas}\ }\textbf {\bibinfo {volume} {19}},\
  \bibinfo {pages} {112105} (\bibinfo {year} {2012})}\BibitemShut {NoStop}%
\bibitem [{\citenamefont {Velikovich}\ and\ \citenamefont
  {Davis}(1995)}]{Velikovich}%
  \BibitemOpen
  \bibfield  {author} {\bibinfo {author} {\bibfnamefont {A.~I.}\ \bibnamefont
  {Velikovich}}\ and\ \bibinfo {author} {\bibfnamefont {J.}~\bibnamefont
  {Davis}},\ }\href {https://doi.org/10.1063/1.871467} {\bibfield  {journal}
  {\bibinfo  {journal} {Phys. Plasmas}\ }\textbf {\bibinfo {volume} {2}},\
  \bibinfo {pages} {4513} (\bibinfo {year} {1995})}\BibitemShut {NoStop}%
\bibitem [{\citenamefont {Knapp}\ \emph {et~al.}(2019)\citenamefont {Knapp},
  \citenamefont {Gomez}, \citenamefont {Hansen}, \citenamefont {Glinsky},
  \citenamefont {Jennings}, \citenamefont {Slutz}, \citenamefont {Harding},
  \citenamefont {Hahn}, \citenamefont {Weis}, \citenamefont {Evans},
  \citenamefont {Martin}, \citenamefont {Harvey-Thompson}, \citenamefont
  {Geissel}, \citenamefont {Smith}, \citenamefont {Ruiz}, \citenamefont
  {Peterson}, \citenamefont {Jones}, \citenamefont {Schwarz}, \citenamefont
  {Rochau}, \citenamefont {Sinars}, \citenamefont {McBride},\ and\
  \citenamefont {Gourdain}}]{Knapp2019}%
  \BibitemOpen
  \bibfield  {author} {\bibinfo {author} {\bibfnamefont {P.~F.}\ \bibnamefont
  {Knapp}}, \bibinfo {author} {\bibfnamefont {M.~R.}\ \bibnamefont {Gomez}},
  \bibinfo {author} {\bibfnamefont {S.~B.}\ \bibnamefont {Hansen}}, \bibinfo
  {author} {\bibfnamefont {M.~E.}\ \bibnamefont {Glinsky}}, \bibinfo {author}
  {\bibfnamefont {C.~A.}\ \bibnamefont {Jennings}}, \bibinfo {author}
  {\bibfnamefont {S.~A.}\ \bibnamefont {Slutz}}, \bibinfo {author}
  {\bibfnamefont {E.~C.}\ \bibnamefont {Harding}}, \bibinfo {author}
  {\bibfnamefont {K.~D.}\ \bibnamefont {Hahn}}, \bibinfo {author}
  {\bibfnamefont {M.~R.}\ \bibnamefont {Weis}}, \bibinfo {author}
  {\bibfnamefont {M.}~\bibnamefont {Evans}}, \bibinfo {author} {\bibfnamefont
  {M.~R.}\ \bibnamefont {Martin}}, \bibinfo {author} {\bibfnamefont {A.~J.}\
  \bibnamefont {Harvey-Thompson}}, \bibinfo {author} {\bibfnamefont
  {M.}~\bibnamefont {Geissel}}, \bibinfo {author} {\bibfnamefont {I.~C.}\
  \bibnamefont {Smith}}, \bibinfo {author} {\bibfnamefont {D.~E.}\ \bibnamefont
  {Ruiz}}, \bibinfo {author} {\bibfnamefont {K.~J.}\ \bibnamefont {Peterson}},
  \bibinfo {author} {\bibfnamefont {B.~M.}\ \bibnamefont {Jones}}, \bibinfo
  {author} {\bibfnamefont {J.}~\bibnamefont {Schwarz}}, \bibinfo {author}
  {\bibfnamefont {G.~A.}\ \bibnamefont {Rochau}}, \bibinfo {author}
  {\bibfnamefont {D.~B.}\ \bibnamefont {Sinars}}, \bibinfo {author}
  {\bibfnamefont {R.~D.}\ \bibnamefont {McBride}},\ and\ \bibinfo {author}
  {\bibfnamefont {P.-A.}\ \bibnamefont {Gourdain}},\ }\href
  {https://doi.org/10.1063/1.5064548} {\bibfield  {journal} {\bibinfo
  {journal} {Phys. Plasmas}\ }\textbf {\bibinfo {volume} {26}},\ \bibinfo
  {pages} {012704} (\bibinfo {year} {2019})}\BibitemShut {NoStop}%
\bibitem [{\citenamefont {Gomez}\ \emph {et~al.}(2019)\citenamefont {Gomez},
  \citenamefont {Slutz}, \citenamefont {Knapp}, \citenamefont {Hahn},
  \citenamefont {Weis}, \citenamefont {Harding}, \citenamefont {Geissel},
  \citenamefont {Fein}, \citenamefont {Glinsky}, \citenamefont {Hansen},
  \citenamefont {Harvey-Thompson}, \citenamefont {Jennings}, \citenamefont
  {Smith}, \citenamefont {Woodbury}, \citenamefont {Ampleford}, \citenamefont
  {Awe}, \citenamefont {Chandler}, \citenamefont {Hess}, \citenamefont
  {Lamppa}, \citenamefont {Myers}, \citenamefont {Ruiz}, \citenamefont
  {Sefkow}, \citenamefont {Schwarz}, \citenamefont {Yager-Elorriaga},
  \citenamefont {Jones}, \citenamefont {Porter}, \citenamefont {Peterson},
  \citenamefont {McBride}, \citenamefont {Rochau},\ and\ \citenamefont
  {Sinars}}]{Gomez2019}%
  \BibitemOpen
  \bibfield  {author} {\bibinfo {author} {\bibfnamefont {M.~R.}\ \bibnamefont
  {Gomez}}, \bibinfo {author} {\bibfnamefont {S.~A.}\ \bibnamefont {Slutz}},
  \bibinfo {author} {\bibfnamefont {P.~F.}\ \bibnamefont {Knapp}}, \bibinfo
  {author} {\bibfnamefont {K.~D.}\ \bibnamefont {Hahn}}, \bibinfo {author}
  {\bibfnamefont {M.~R.}\ \bibnamefont {Weis}}, \bibinfo {author}
  {\bibfnamefont {E.~C.}\ \bibnamefont {Harding}}, \bibinfo {author}
  {\bibfnamefont {M.}~\bibnamefont {Geissel}}, \bibinfo {author} {\bibfnamefont
  {J.~R.}\ \bibnamefont {Fein}}, \bibinfo {author} {\bibfnamefont {M.~E.}\
  \bibnamefont {Glinsky}}, \bibinfo {author} {\bibfnamefont {S.~B.}\
  \bibnamefont {Hansen}}, \bibinfo {author} {\bibfnamefont {A.~J.}\
  \bibnamefont {Harvey-Thompson}}, \bibinfo {author} {\bibfnamefont {C.~A.}\
  \bibnamefont {Jennings}}, \bibinfo {author} {\bibfnamefont {I.~C.}\
  \bibnamefont {Smith}}, \bibinfo {author} {\bibfnamefont {D.}~\bibnamefont
  {Woodbury}}, \bibinfo {author} {\bibfnamefont {D.~J.}\ \bibnamefont
  {Ampleford}}, \bibinfo {author} {\bibfnamefont {T.~J.}\ \bibnamefont {Awe}},
  \bibinfo {author} {\bibfnamefont {G.~A.}\ \bibnamefont {Chandler}}, \bibinfo
  {author} {\bibfnamefont {M.~H.}\ \bibnamefont {Hess}}, \bibinfo {author}
  {\bibfnamefont {D.~C.}\ \bibnamefont {Lamppa}}, \bibinfo {author}
  {\bibfnamefont {C.~E.}\ \bibnamefont {Myers}}, \bibinfo {author}
  {\bibfnamefont {C.~L.}\ \bibnamefont {Ruiz}}, \bibinfo {author}
  {\bibfnamefont {A.~B.}\ \bibnamefont {Sefkow}}, \bibinfo {author}
  {\bibfnamefont {J.}~\bibnamefont {Schwarz}}, \bibinfo {author} {\bibfnamefont
  {D.~A.}\ \bibnamefont {Yager-Elorriaga}}, \bibinfo {author} {\bibfnamefont
  {B.}~\bibnamefont {Jones}}, \bibinfo {author} {\bibfnamefont {J.~L.}\
  \bibnamefont {Porter}}, \bibinfo {author} {\bibfnamefont {K.~J.}\
  \bibnamefont {Peterson}}, \bibinfo {author} {\bibfnamefont {R.~D.}\
  \bibnamefont {McBride}}, \bibinfo {author} {\bibfnamefont {G.~A.}\
  \bibnamefont {Rochau}},\ and\ \bibinfo {author} {\bibfnamefont {D.~B.}\
  \bibnamefont {Sinars}},\ }\href {https://doi.org/10.1109/TPS.2019.2893517}
  {\bibfield  {journal} {\bibinfo  {journal} {IEEE Plasma Sci.}\ }\textbf
  {\bibinfo {volume} {47}},\ \bibinfo {pages} {2081} (\bibinfo {year}
  {2019})}\BibitemShut {NoStop}%
\bibitem [{\citenamefont {Slutz}\ and\ \citenamefont
  {Vesey}(2012)}]{Slutz2012}%
  \BibitemOpen
  \bibfield  {author} {\bibinfo {author} {\bibfnamefont {S.~A.}\ \bibnamefont
  {Slutz}}\ and\ \bibinfo {author} {\bibfnamefont {R.~A.}\ \bibnamefont
  {Vesey}},\ }\href {https://doi.org/10.1103/PhysRevLett.108.025003} {\bibfield
   {journal} {\bibinfo  {journal} {Phys. Rev. Lett.}\ }\textbf {\bibinfo
  {volume} {108}},\ \bibinfo {pages} {025003} (\bibinfo {year}
  {2012})}\BibitemShut {NoStop}%
\bibitem [{\citenamefont {Gomez}\ \emph {et~al.}(2014)\citenamefont {Gomez},
  \citenamefont {Slutz}, \citenamefont {Sefkow}, \citenamefont {Sinars},
  \citenamefont {Hahn}, \citenamefont {Hansen}, \citenamefont {Harding},
  \citenamefont {Knapp}, \citenamefont {Schmit}, \citenamefont {Jennings},
  \citenamefont {Awe}, \citenamefont {Geissel}, \citenamefont {Rovang},
  \citenamefont {Chandler}, \citenamefont {Cooper}, \citenamefont {Cuneo},
  \citenamefont {Harvey-Thompson}, \citenamefont {Herrmann}, \citenamefont
  {Hess}, \citenamefont {Johns}, \citenamefont {Lamppa}, \citenamefont
  {Martin}, \citenamefont {McBride}, \citenamefont {Peterson}, \citenamefont
  {Porter}, \citenamefont {Robertson}, \citenamefont {Rochau}, \citenamefont
  {Ruiz}, \citenamefont {Savage}, \citenamefont {Smith}, \citenamefont
  {Stygar},\ and\ \citenamefont {Vesey}}]{Gomez2014}%
  \BibitemOpen
  \bibfield  {author} {\bibinfo {author} {\bibfnamefont {M.~R.}\ \bibnamefont
  {Gomez}}, \bibinfo {author} {\bibfnamefont {S.~A.}\ \bibnamefont {Slutz}},
  \bibinfo {author} {\bibfnamefont {A.~B.}\ \bibnamefont {Sefkow}}, \bibinfo
  {author} {\bibfnamefont {D.~B.}\ \bibnamefont {Sinars}}, \bibinfo {author}
  {\bibfnamefont {K.~D.}\ \bibnamefont {Hahn}}, \bibinfo {author}
  {\bibfnamefont {S.~B.}\ \bibnamefont {Hansen}}, \bibinfo {author}
  {\bibfnamefont {E.~C.}\ \bibnamefont {Harding}}, \bibinfo {author}
  {\bibfnamefont {P.~F.}\ \bibnamefont {Knapp}}, \bibinfo {author}
  {\bibfnamefont {P.~F.}\ \bibnamefont {Schmit}}, \bibinfo {author}
  {\bibfnamefont {C.~A.}\ \bibnamefont {Jennings}}, \bibinfo {author}
  {\bibfnamefont {T.~J.}\ \bibnamefont {Awe}}, \bibinfo {author} {\bibfnamefont
  {M.}~\bibnamefont {Geissel}}, \bibinfo {author} {\bibfnamefont {D.~C.}\
  \bibnamefont {Rovang}}, \bibinfo {author} {\bibfnamefont {G.~A.}\
  \bibnamefont {Chandler}}, \bibinfo {author} {\bibfnamefont {G.~W.}\
  \bibnamefont {Cooper}}, \bibinfo {author} {\bibfnamefont {M.~E.}\
  \bibnamefont {Cuneo}}, \bibinfo {author} {\bibfnamefont {A.~J.}\ \bibnamefont
  {Harvey-Thompson}}, \bibinfo {author} {\bibfnamefont {M.~C.}\ \bibnamefont
  {Herrmann}}, \bibinfo {author} {\bibfnamefont {M.~H.}\ \bibnamefont {Hess}},
  \bibinfo {author} {\bibfnamefont {O.}~\bibnamefont {Johns}}, \bibinfo
  {author} {\bibfnamefont {D.~C.}\ \bibnamefont {Lamppa}}, \bibinfo {author}
  {\bibfnamefont {M.~R.}\ \bibnamefont {Martin}}, \bibinfo {author}
  {\bibfnamefont {R.~D.}\ \bibnamefont {McBride}}, \bibinfo {author}
  {\bibfnamefont {K.~J.}\ \bibnamefont {Peterson}}, \bibinfo {author}
  {\bibfnamefont {J.~L.}\ \bibnamefont {Porter}}, \bibinfo {author}
  {\bibfnamefont {G.~K.}\ \bibnamefont {Robertson}}, \bibinfo {author}
  {\bibfnamefont {G.~A.}\ \bibnamefont {Rochau}}, \bibinfo {author}
  {\bibfnamefont {C.~L.}\ \bibnamefont {Ruiz}}, \bibinfo {author}
  {\bibfnamefont {M.~E.}\ \bibnamefont {Savage}}, \bibinfo {author}
  {\bibfnamefont {I.~C.}\ \bibnamefont {Smith}}, \bibinfo {author}
  {\bibfnamefont {W.~A.}\ \bibnamefont {Stygar}},\ and\ \bibinfo {author}
  {\bibfnamefont {R.~A.}\ \bibnamefont {Vesey}},\ }\href
  {https://doi.org/10.1103/PhysRevLett.113.155003} {\bibfield  {journal}
  {\bibinfo  {journal} {Phys. Rev. Lett.}\ }\textbf {\bibinfo {volume} {113}},\
  \bibinfo {pages} {155003} (\bibinfo {year} {2014})}\BibitemShut {NoStop}%
\bibitem [{\citenamefont {Bekhtenev}\ \emph {et~al.}(1980)\citenamefont
  {Bekhtenev}, \citenamefont {Volosov}, \citenamefont {Pal'Chikov},
  \citenamefont {Pekker},\ and\ \citenamefont {Yudin}}]{bekhtenev1980problems}%
  \BibitemOpen
  \bibfield  {author} {\bibinfo {author} {\bibfnamefont {A.}~\bibnamefont
  {Bekhtenev}}, \bibinfo {author} {\bibfnamefont {V.}~\bibnamefont {Volosov}},
  \bibinfo {author} {\bibfnamefont {V.}~\bibnamefont {Pal'Chikov}}, \bibinfo
  {author} {\bibfnamefont {M.}~\bibnamefont {Pekker}},\ and\ \bibinfo {author}
  {\bibfnamefont {Y.~N.}\ \bibnamefont {Yudin}},\ }\href@noop {} {\bibfield
  {journal} {\bibinfo  {journal} {Nuclear Fusion}\ }\textbf {\bibinfo {volume}
  {20}},\ \bibinfo {pages} {579} (\bibinfo {year} {1980})}\BibitemShut
  {NoStop}%
\bibitem [{\citenamefont {Ellis}\ \emph {et~al.}(2001)\citenamefont {Ellis},
  \citenamefont {Hassam}, \citenamefont {Messer},\ and\ \citenamefont
  {Osborn}}]{Ellis2001}%
  \BibitemOpen
  \bibfield  {author} {\bibinfo {author} {\bibfnamefont {R.~F.}\ \bibnamefont
  {Ellis}}, \bibinfo {author} {\bibfnamefont {A.~B.}\ \bibnamefont {Hassam}},
  \bibinfo {author} {\bibfnamefont {S.}~\bibnamefont {Messer}},\ and\ \bibinfo
  {author} {\bibfnamefont {B.~R.}\ \bibnamefont {Osborn}},\ }\href
  {https://doi.org/10.1063/1.1350957} {\bibfield  {journal} {\bibinfo
  {journal} {Physics of Plasmas}\ }\textbf {\bibinfo {volume} {8}},\ \bibinfo
  {pages} {2057} (\bibinfo {year} {2001})}\BibitemShut {NoStop}%
\bibitem [{\citenamefont {Fetterman}\ and\ \citenamefont
  {Fisch}(2008)}]{Fetterman2008}%
  \BibitemOpen
  \bibfield  {author} {\bibinfo {author} {\bibfnamefont {A.~J.}\ \bibnamefont
  {Fetterman}}\ and\ \bibinfo {author} {\bibfnamefont {N.~J.}\ \bibnamefont
  {Fisch}},\ }\href {https://doi.org/10.1103/PhysRevLett.101.205003} {\bibfield
   {journal} {\bibinfo  {journal} {Phys. Rev. Lett.}\ }\textbf {\bibinfo
  {volume} {101}},\ \bibinfo {pages} {205003} (\bibinfo {year}
  {2008})}\BibitemShut {NoStop}%
\bibitem [{\citenamefont {Kolmes}\ \emph
  {et~al.}(2018{\natexlab{a}})\citenamefont {Kolmes}, \citenamefont {Ochs},\
  and\ \citenamefont {Fisch}}]{Kolmes2018}%
  \BibitemOpen
  \bibfield  {author} {\bibinfo {author} {\bibfnamefont {E.~J.}\ \bibnamefont
  {Kolmes}}, \bibinfo {author} {\bibfnamefont {I.~E.}\ \bibnamefont {Ochs}},\
  and\ \bibinfo {author} {\bibfnamefont {N.~J.}\ \bibnamefont {Fisch}},\ }\href
  {https://doi.org/10.1063/1.5023931} {\bibfield  {journal} {\bibinfo
  {journal} {Phys. Plasmas}\ }\textbf {\bibinfo {volume} {25}},\ \bibinfo
  {pages} {032508} (\bibinfo {year} {2018}{\natexlab{a}})}\BibitemShut
  {NoStop}%
\bibitem [{\citenamefont {Rax}\ \emph {et~al.}(2017)\citenamefont {Rax},
  \citenamefont {Gueroult},\ and\ \citenamefont {Fisch}}]{Rax2017}%
  \BibitemOpen
  \bibfield  {author} {\bibinfo {author} {\bibfnamefont {J.-M.}\ \bibnamefont
  {Rax}}, \bibinfo {author} {\bibfnamefont {R.}~\bibnamefont {Gueroult}},\ and\
  \bibinfo {author} {\bibfnamefont {N.~J.}\ \bibnamefont {Fisch}},\ }\href
  {https://doi.org/10.1063/1.4977919} {\bibfield  {journal} {\bibinfo
  {journal} {Phys. Plasmas}\ }\textbf {\bibinfo {volume} {24}},\ \bibinfo
  {pages} {032504} (\bibinfo {year} {2017})}\BibitemShut {NoStop}%
\bibitem [{\citenamefont {Bonnevier}(1966)}]{Bonnevier1966}%
  \BibitemOpen
  \bibfield  {author} {\bibinfo {author} {\bibfnamefont {B.}~\bibnamefont
  {Bonnevier}},\ }\href@noop {} {\bibfield  {journal} {\bibinfo  {journal}
  {Ark. Fys.}\ }\textbf {\bibinfo {volume} {33}},\ \bibinfo {pages} {255}
  (\bibinfo {year} {1966})}\BibitemShut {NoStop}%
\bibitem [{\citenamefont {Lehnert}(1971)}]{Lehnert1971}%
  \BibitemOpen
  \bibfield  {author} {\bibinfo {author} {\bibfnamefont {B.}~\bibnamefont
  {Lehnert}},\ }\href {https://doi.org/10.1088/0029-5515/11/5/010} {\bibfield
  {journal} {\bibinfo  {journal} {Nucl. Fusion}\ }\textbf {\bibinfo {volume}
  {11}},\ \bibinfo {pages} {485} (\bibinfo {year} {1971})}\BibitemShut
  {NoStop}%
\bibitem [{\citenamefont {Hellsten}(1977)}]{Hellsten1977}%
  \BibitemOpen
  \bibfield  {author} {\bibinfo {author} {\bibfnamefont {T.}~\bibnamefont
  {Hellsten}},\ }\href {https://doi.org/10.1016/0029-554X(77)90572-9}
  {\bibfield  {journal} {\bibinfo  {journal} {Nucl. Instr. and Meth.}\ }\textbf
  {\bibinfo {volume} {145}},\ \bibinfo {pages} {425} (\bibinfo {year}
  {1977})}\BibitemShut {NoStop}%
\bibitem [{\citenamefont {Krishnan}(1983)}]{Krishnan1983}%
  \BibitemOpen
  \bibfield  {author} {\bibinfo {author} {\bibfnamefont {M.}~\bibnamefont
  {Krishnan}},\ }\href {https://doi.org/10.1063/1.864460} {\bibfield  {journal}
  {\bibinfo  {journal} {Phys. Fluids}\ }\textbf {\bibinfo {volume} {26}},\
  \bibinfo {pages} {2676} (\bibinfo {year} {1983})}\BibitemShut {NoStop}%
\bibitem [{\citenamefont {Geva}\ \emph {et~al.}(1984)\citenamefont {Geva},
  \citenamefont {Krishnan},\ and\ \citenamefont {Hirshfield}}]{Geva1984}%
  \BibitemOpen
  \bibfield  {author} {\bibinfo {author} {\bibfnamefont {M.}~\bibnamefont
  {Geva}}, \bibinfo {author} {\bibfnamefont {M.}~\bibnamefont {Krishnan}},\
  and\ \bibinfo {author} {\bibfnamefont {J.~L.}\ \bibnamefont {Hirshfield}},\
  }\href {https://doi.org/10.1063/1.334139} {\bibfield  {journal} {\bibinfo
  {journal} {J. Appl. Phys.}\ }\textbf {\bibinfo {volume} {56}},\ \bibinfo
  {pages} {1398} (\bibinfo {year} {1984})}\BibitemShut {NoStop}%
\bibitem [{\citenamefont {Dolgolenko}\ and\ \citenamefont {{\relax Yu. A.
  Muromkin}}(2017)}]{Dolgolenko2017}%
  \BibitemOpen
  \bibfield  {author} {\bibinfo {author} {\bibfnamefont {D.~A.}\ \bibnamefont
  {Dolgolenko}}\ and\ \bibinfo {author} {\bibnamefont {{\relax Yu. A.
  Muromkin}}},\ }\href {https://doi.org/10.3367/UFNe.2016.12.038016} {\bibfield
   {journal} {\bibinfo  {journal} {Phys.-Usp.}\ }\textbf {\bibinfo {volume}
  {60}},\ \bibinfo {pages} {994} (\bibinfo {year} {2017})}\BibitemShut
  {NoStop}%
\bibitem [{\citenamefont {Ochs}\ \emph {et~al.}(2017)\citenamefont {Ochs},
  \citenamefont {Gueroult}, \citenamefont {Fisch},\ and\ \citenamefont
  {Zweben}}]{Ochs2017iii}%
  \BibitemOpen
  \bibfield  {author} {\bibinfo {author} {\bibfnamefont {I.~E.}\ \bibnamefont
  {Ochs}}, \bibinfo {author} {\bibfnamefont {R.}~\bibnamefont {Gueroult}},
  \bibinfo {author} {\bibfnamefont {N.~J.}\ \bibnamefont {Fisch}},\ and\
  \bibinfo {author} {\bibfnamefont {S.~J.}\ \bibnamefont {Zweben}},\ }\href
  {https://doi.org/10.1063/1.4978949} {\bibfield  {journal} {\bibinfo
  {journal} {Phys. Plasmas}\ }\textbf {\bibinfo {volume} {24}},\ \bibinfo
  {pages} {043503} (\bibinfo {year} {2017})}\BibitemShut {NoStop}%
\bibitem [{\citenamefont {Yuferov}\ \emph {et~al.}(2018)\citenamefont
  {Yuferov}, \citenamefont {Katrechko}, \citenamefont {Ilichova}, \citenamefont
  {Shariy}, \citenamefont {Svichkar}, \citenamefont {Shvets}, \citenamefont
  {Mufel},\ and\ \citenamefont {Bobrov}}]{Yuferov2018}%
  \BibitemOpen
  \bibfield  {author} {\bibinfo {author} {\bibfnamefont {V.~B.}\ \bibnamefont
  {Yuferov}}, \bibinfo {author} {\bibfnamefont {S.~V.}\ \bibnamefont
  {Katrechko}}, \bibinfo {author} {\bibfnamefont {V.~O.}\ \bibnamefont
  {Ilichova}}, \bibinfo {author} {\bibfnamefont {S.~V.}\ \bibnamefont
  {Shariy}}, \bibinfo {author} {\bibfnamefont {A.~S.}\ \bibnamefont
  {Svichkar}}, \bibinfo {author} {\bibfnamefont {M.~O.}\ \bibnamefont
  {Shvets}}, \bibinfo {author} {\bibfnamefont {E.~V.}\ \bibnamefont {Mufel}},\
  and\ \bibinfo {author} {\bibfnamefont {A.~G.}\ \bibnamefont {Bobrov}},\
  }\href@noop {} {\bibfield  {journal} {\bibinfo  {journal} {Prob. Atomic Sci.
  and Tech.}\ }\textbf {\bibinfo {volume} {113}},\ \bibinfo {pages} {118}
  (\bibinfo {year} {2018})}\BibitemShut {NoStop}%
\bibitem [{\citenamefont {Gueroult}\ \emph
  {et~al.}(2018{\natexlab{a}})\citenamefont {Gueroult}, \citenamefont {Rax},
  \citenamefont {Zweben},\ and\ \citenamefont {Fisch}}]{Gueroult2018}%
  \BibitemOpen
  \bibfield  {author} {\bibinfo {author} {\bibfnamefont {R.}~\bibnamefont
  {Gueroult}}, \bibinfo {author} {\bibfnamefont {J.-M.}\ \bibnamefont {Rax}},
  \bibinfo {author} {\bibfnamefont {S.~J.}\ \bibnamefont {Zweben}},\ and\
  \bibinfo {author} {\bibfnamefont {N.~J.}\ \bibnamefont {Fisch}},\ }\href
  {https://doi.org/10.1088/1361-6587/aa8be5} {\bibfield  {journal} {\bibinfo
  {journal} {Plasma Phys. and Control. Fusion}\ }\textbf {\bibinfo {volume}
  {60}},\ \bibinfo {pages} {014018} (\bibinfo {year}
  {2018}{\natexlab{a}})}\BibitemShut {NoStop}%
\bibitem [{\citenamefont {Fetterman}\ and\ \citenamefont
  {Fisch}(2011)}]{Fetterman2011b}%
  \BibitemOpen
  \bibfield  {author} {\bibinfo {author} {\bibfnamefont {A.~J.}\ \bibnamefont
  {Fetterman}}\ and\ \bibinfo {author} {\bibfnamefont {N.~J.}\ \bibnamefont
  {Fisch}},\ }\href {https://doi.org/10.1063/1.3631793} {\bibfield  {journal}
  {\bibinfo  {journal} {Phys. Plasmas}\ }\textbf {\bibinfo {volume} {18}},\
  \bibinfo {pages} {094503} (\bibinfo {year} {2011})}\BibitemShut {NoStop}%
\bibitem [{\citenamefont {Ohkawa}\ and\ \citenamefont
  {Miller}(2002)}]{Ohkawa2002}%
  \BibitemOpen
  \bibfield  {author} {\bibinfo {author} {\bibfnamefont {T.}~\bibnamefont
  {Ohkawa}}\ and\ \bibinfo {author} {\bibfnamefont {R.~L.}\ \bibnamefont
  {Miller}},\ }\href {https://doi.org/10.1063/1.1523930} {\bibfield  {journal}
  {\bibinfo  {journal} {Phys. Plasmas}\ }\textbf {\bibinfo {volume} {9}},\
  \bibinfo {pages} {5116} (\bibinfo {year} {2002})}\BibitemShut {NoStop}%
\bibitem [{\citenamefont {O'Neil}(1981)}]{ONeil1981}%
  \BibitemOpen
  \bibfield  {author} {\bibinfo {author} {\bibfnamefont {T.~M.}\ \bibnamefont
  {O'Neil}},\ }\href {https://doi.org/10.1063/1.863565} {\bibfield  {journal}
  {\bibinfo  {journal} {Phys. Fluids}\ }\textbf {\bibinfo {volume} {24}},\
  \bibinfo {pages} {1447} (\bibinfo {year} {1981})}\BibitemShut {NoStop}%
\bibitem [{\citenamefont {Rax}\ and\ \citenamefont {Gueroult}(2016)}]{Rax2016}%
  \BibitemOpen
  \bibfield  {author} {\bibinfo {author} {\bibfnamefont {J.-M.}\ \bibnamefont
  {Rax}}\ and\ \bibinfo {author} {\bibfnamefont {R.}~\bibnamefont {Gueroult}},\
  }\href {https://doi.org/10.1017/S0022377816000878} {\bibfield  {journal}
  {\bibinfo  {journal} {{J. Plasma Phys.}}\ }\textbf {\bibinfo {volume}
  {{82}}},\ \bibinfo {pages} {{595820504}} (\bibinfo {year}
  {{2016}})}\BibitemShut {NoStop}%
\bibitem [{\citenamefont {Gueroult}\ \emph
  {et~al.}(2018{\natexlab{b}})\citenamefont {Gueroult}, \citenamefont {Rax},\
  and\ \citenamefont {Fisch}}]{Gueroult2018ii}%
  \BibitemOpen
  \bibfield  {author} {\bibinfo {author} {\bibfnamefont {R.}~\bibnamefont
  {Gueroult}}, \bibinfo {author} {\bibfnamefont {J.-M.}\ \bibnamefont {Rax}},\
  and\ \bibinfo {author} {\bibfnamefont {N.~J.}\ \bibnamefont {Fisch}},\ }\href
  {https://doi.org/10.1016/j.jclepro.2018.02.066} {\bibfield  {journal}
  {\bibinfo  {journal} {J. Clean. Prod.}\ }\textbf {\bibinfo {volume} {182}},\
  \bibinfo {pages} {1060} (\bibinfo {year} {2018}{\natexlab{b}})}\BibitemShut
  {NoStop}%
\bibitem [{\citenamefont {Braginskii}(1965)}]{Braginskii1965}%
  \BibitemOpen
  \bibfield  {author} {\bibinfo {author} {\bibfnamefont {S.~I.}\ \bibnamefont
  {Braginskii}},\ }\bibinfo {title} {Transport processes in a plasma},\ in\
  \href@noop {} {\emph {\bibinfo {booktitle} {Reviews of Plasma Physics}}},\
  Vol.~\bibinfo {volume} {1},\ \bibinfo {editor} {edited by\ \bibinfo {editor}
  {\bibfnamefont {M.~A.}\ \bibnamefont {Leontovich}}}\ (\bibinfo  {publisher}
  {Consultants Bureau},\ \bibinfo {address} {New York},\ \bibinfo {year}
  {1965})\ p.\ \bibinfo {pages} {205}\BibitemShut {NoStop}%
\bibitem [{\citenamefont {Zhdanov}(2002)}]{Zhdanov_2002}%
  \BibitemOpen
  \bibfield  {author} {\bibinfo {author} {\bibfnamefont {V.~M.}\ \bibnamefont
  {Zhdanov}},\ }\href {https://doi.org/10.1088/0741-3335/44/10/701} {\bibfield
  {journal} {\bibinfo  {journal} {Plasma Physics and Controlled Fusion}\
  }\textbf {\bibinfo {volume} {44}},\ \bibinfo {pages} {2283} (\bibinfo {year}
  {2002})}\BibitemShut {NoStop}%
\bibitem [{\citenamefont {{Litvinova Mitra}}\ \emph {et~al.}(2021)\citenamefont
  {{Litvinova Mitra}}, \citenamefont {Kolmes}, \citenamefont {Ochs},
  \citenamefont {Mlodik}, \citenamefont {Rubin},\ and\ \citenamefont
  {Fisch}}]{Litvinova2021}%
  \BibitemOpen
  \bibfield  {author} {\bibinfo {author} {\bibfnamefont {E.}~\bibnamefont
  {{Litvinova Mitra}}}, \bibinfo {author} {\bibfnamefont {E.}~\bibnamefont
  {Kolmes}}, \bibinfo {author} {\bibfnamefont {I.}~\bibnamefont {Ochs}},
  \bibinfo {author} {\bibfnamefont {M.}~\bibnamefont {Mlodik}}, \bibinfo
  {author} {\bibfnamefont {T.}~\bibnamefont {Rubin}},\ and\ \bibinfo {author}
  {\bibfnamefont {N.}~\bibnamefont {Fisch}},\ }\href
  {https://doi.org/https://doi.org/10.1016/j.physleta.2021.127284} {\bibfield
  {journal} {\bibinfo  {journal} {Physics Letters A}\ }\textbf {\bibinfo
  {volume} {398}},\ \bibinfo {pages} {127284} (\bibinfo {year}
  {2021})}\BibitemShut {NoStop}%
\bibitem [{\citenamefont {Kolmes}\ \emph {et~al.}(2019)\citenamefont {Kolmes},
  \citenamefont {Ochs}, \citenamefont {Mlodik}, \citenamefont {Rax},
  \citenamefont {Gueroult},\ and\ \citenamefont {Fisch}}]{Radial_Current}%
  \BibitemOpen
  \bibfield  {author} {\bibinfo {author} {\bibfnamefont {E.~J.}\ \bibnamefont
  {Kolmes}}, \bibinfo {author} {\bibfnamefont {I.~E.}\ \bibnamefont {Ochs}},
  \bibinfo {author} {\bibfnamefont {M.~E.}\ \bibnamefont {Mlodik}}, \bibinfo
  {author} {\bibfnamefont {J.-M.}\ \bibnamefont {Rax}}, \bibinfo {author}
  {\bibfnamefont {R.}~\bibnamefont {Gueroult}},\ and\ \bibinfo {author}
  {\bibfnamefont {N.~J.}\ \bibnamefont {Fisch}},\ }\href
  {https://doi.org/10.1063/1.5115788} {\bibfield  {journal} {\bibinfo
  {journal} {Physics of Plasmas}\ }\textbf {\bibinfo {volume} {26}},\ \bibinfo
  {pages} {082309} (\bibinfo {year} {2019})}\BibitemShut {NoStop}%
\bibitem [{\citenamefont {Kolmes}\ \emph
  {et~al.}(2018{\natexlab{b}})\citenamefont {Kolmes}, \citenamefont {Ochs},\
  and\ \citenamefont {Fisch}}]{Strategies}%
  \BibitemOpen
  \bibfield  {author} {\bibinfo {author} {\bibfnamefont {E.~J.}\ \bibnamefont
  {Kolmes}}, \bibinfo {author} {\bibfnamefont {I.~E.}\ \bibnamefont {Ochs}},\
  and\ \bibinfo {author} {\bibfnamefont {N.~J.}\ \bibnamefont {Fisch}},\ }\href
  {https://doi.org/10.1063/1.5023931} {\bibfield  {journal} {\bibinfo
  {journal} {Physics of Plasmas}\ }\textbf {\bibinfo {volume} {25}},\ \bibinfo
  {pages} {032508} (\bibinfo {year} {2018}{\natexlab{b}})}\BibitemShut
  {NoStop}%
\bibitem [{\citenamefont {Mlodik}\ \emph {et~al.}(2021)\citenamefont {Mlodik},
  \citenamefont {Kolmes}, \citenamefont {Ochs},\ and\ \citenamefont
  {Fisch}}]{Mlodik2021}%
  \BibitemOpen
  \bibfield  {author} {\bibinfo {author} {\bibfnamefont {M.~E.}\ \bibnamefont
  {Mlodik}}, \bibinfo {author} {\bibfnamefont {E.~J.}\ \bibnamefont {Kolmes}},
  \bibinfo {author} {\bibfnamefont {I.~E.}\ \bibnamefont {Ochs}},\ and\
  \bibinfo {author} {\bibfnamefont {N.~J.}\ \bibnamefont {Fisch}},\ }\href
  {https://doi.org/10.1063/5.0046603} {\bibfield  {journal} {\bibinfo
  {journal} {Physics of Plasmas}\ }\textbf {\bibinfo {volume} {28}},\ \bibinfo
  {pages} {052702} (\bibinfo {year} {2021})}\BibitemShut {NoStop}%
\bibitem [{\citenamefont {Kolmes}\ \emph
  {et~al.}(2021{\natexlab{a}})\citenamefont {Kolmes}, \citenamefont {Ochs},
  \citenamefont {Mlodik},\ and\ \citenamefont {Fisch}}]{HotIonMode}%
  \BibitemOpen
  \bibfield  {author} {\bibinfo {author} {\bibfnamefont {E.~J.}\ \bibnamefont
  {Kolmes}}, \bibinfo {author} {\bibfnamefont {I.~E.}\ \bibnamefont {Ochs}},
  \bibinfo {author} {\bibfnamefont {M.~E.}\ \bibnamefont {Mlodik}},\ and\
  \bibinfo {author} {\bibfnamefont {N.~J.}\ \bibnamefont {Fisch}},\ }\href
  {https://doi.org/10.1103/PhysRevE.104.015209} {\bibfield  {journal} {\bibinfo
   {journal} {Phys. Rev. E}\ }\textbf {\bibinfo {volume} {104}},\ \bibinfo
  {pages} {015209} (\bibinfo {year} {2021}{\natexlab{a}})}\BibitemShut
  {NoStop}%
\bibitem [{\citenamefont {Mlodik}\ \emph {et~al.}(2020)\citenamefont {Mlodik},
  \citenamefont {Kolmes}, \citenamefont {Ochs},\ and\ \citenamefont
  {Fisch}}]{Mlodik2020}%
  \BibitemOpen
  \bibfield  {author} {\bibinfo {author} {\bibfnamefont {M.~E.}\ \bibnamefont
  {Mlodik}}, \bibinfo {author} {\bibfnamefont {E.~J.}\ \bibnamefont {Kolmes}},
  \bibinfo {author} {\bibfnamefont {I.~E.}\ \bibnamefont {Ochs}},\ and\
  \bibinfo {author} {\bibfnamefont {N.~J.}\ \bibnamefont {Fisch}},\ }\href
  {https://doi.org/10.1103/PhysRevE.102.013212} {\bibfield  {journal} {\bibinfo
   {journal} {Phys. Rev. E}\ }\textbf {\bibinfo {volume} {102}},\ \bibinfo
  {pages} {013212} (\bibinfo {year} {2020})}\BibitemShut {NoStop}%
\bibitem [{\citenamefont {Ricci}\ \emph {et~al.}(2012)\citenamefont {Ricci},
  \citenamefont {Halpern}, \citenamefont {Jolliet}, \citenamefont {Loizu},
  \citenamefont {Mosetto}, \citenamefont {Fasoli}, \citenamefont {Furno},\ and\
  \citenamefont {Theiler}}]{ricci2012}%
  \BibitemOpen
  \bibfield  {author} {\bibinfo {author} {\bibfnamefont {P.}~\bibnamefont
  {Ricci}}, \bibinfo {author} {\bibfnamefont {F.}~\bibnamefont {Halpern}},
  \bibinfo {author} {\bibfnamefont {S.}~\bibnamefont {Jolliet}}, \bibinfo
  {author} {\bibfnamefont {J.}~\bibnamefont {Loizu}}, \bibinfo {author}
  {\bibfnamefont {A.}~\bibnamefont {Mosetto}}, \bibinfo {author} {\bibfnamefont
  {A.}~\bibnamefont {Fasoli}}, \bibinfo {author} {\bibfnamefont
  {I.}~\bibnamefont {Furno}},\ and\ \bibinfo {author} {\bibfnamefont
  {C.}~\bibnamefont {Theiler}},\ }\href@noop {} {\bibfield  {journal} {\bibinfo
   {journal} {Plasma Physics and Controlled Fusion}\ }\textbf {\bibinfo
  {volume} {54}},\ \bibinfo {pages} {124047} (\bibinfo {year}
  {2012})}\BibitemShut {NoStop}%
\bibitem [{\citenamefont {Leake}\ \emph {et~al.}(2014)\citenamefont {Leake},
  \citenamefont {DeVore}, \citenamefont {Thayer}, \citenamefont {Burns},
  \citenamefont {Crowley}, \citenamefont {Gilbert}, \citenamefont {Huba},
  \citenamefont {Krall}, \citenamefont {Linton}, \citenamefont {Lukin} \emph
  {et~al.}}]{leake2014}%
  \BibitemOpen
  \bibfield  {author} {\bibinfo {author} {\bibfnamefont {J.}~\bibnamefont
  {Leake}}, \bibinfo {author} {\bibfnamefont {C.}~\bibnamefont {DeVore}},
  \bibinfo {author} {\bibfnamefont {J.}~\bibnamefont {Thayer}}, \bibinfo
  {author} {\bibfnamefont {A.}~\bibnamefont {Burns}}, \bibinfo {author}
  {\bibfnamefont {G.}~\bibnamefont {Crowley}}, \bibinfo {author} {\bibfnamefont
  {H.}~\bibnamefont {Gilbert}}, \bibinfo {author} {\bibfnamefont
  {J.}~\bibnamefont {Huba}}, \bibinfo {author} {\bibfnamefont {J.}~\bibnamefont
  {Krall}}, \bibinfo {author} {\bibfnamefont {M.}~\bibnamefont {Linton}},
  \bibinfo {author} {\bibfnamefont {V.}~\bibnamefont {Lukin}}, \emph {et~al.},\
  }\href@noop {} {\bibfield  {journal} {\bibinfo  {journal} {Space Science
  Reviews}\ }\textbf {\bibinfo {volume} {184}},\ \bibinfo {pages} {107}
  (\bibinfo {year} {2014})}\BibitemShut {NoStop}%
\bibitem [{\citenamefont {Laguna}\ \emph {et~al.}(2016)\citenamefont {Laguna},
  \citenamefont {Lani}, \citenamefont {Deconinck}, \citenamefont {Mansour},\
  and\ \citenamefont {Poedts}}]{laguna2016}%
  \BibitemOpen
  \bibfield  {author} {\bibinfo {author} {\bibfnamefont {A.~A.}\ \bibnamefont
  {Laguna}}, \bibinfo {author} {\bibfnamefont {A.}~\bibnamefont {Lani}},
  \bibinfo {author} {\bibfnamefont {H.}~\bibnamefont {Deconinck}}, \bibinfo
  {author} {\bibfnamefont {N.}~\bibnamefont {Mansour}},\ and\ \bibinfo {author}
  {\bibfnamefont {S.}~\bibnamefont {Poedts}},\ }\href@noop {} {\bibfield
  {journal} {\bibinfo  {journal} {Journal of Computational Physics}\ }\textbf
  {\bibinfo {volume} {318}},\ \bibinfo {pages} {252} (\bibinfo {year}
  {2016})}\BibitemShut {NoStop}%
\bibitem [{\citenamefont {Rognlien}\ \emph {et~al.}(1994)\citenamefont
  {Rognlien}, \citenamefont {Brown}, \citenamefont {Campbell}, \citenamefont
  {Kaiser}, \citenamefont {Knoll}, \citenamefont {McHugh}, \citenamefont
  {Porter}, \citenamefont {Rensink},\ and\ \citenamefont
  {Smith}}]{rognlien19942}%
  \BibitemOpen
  \bibfield  {author} {\bibinfo {author} {\bibfnamefont {T.}~\bibnamefont
  {Rognlien}}, \bibinfo {author} {\bibfnamefont {P.}~\bibnamefont {Brown}},
  \bibinfo {author} {\bibfnamefont {R.}~\bibnamefont {Campbell}}, \bibinfo
  {author} {\bibfnamefont {T.}~\bibnamefont {Kaiser}}, \bibinfo {author}
  {\bibfnamefont {D.}~\bibnamefont {Knoll}}, \bibinfo {author} {\bibfnamefont
  {P.}~\bibnamefont {McHugh}}, \bibinfo {author} {\bibfnamefont
  {G.}~\bibnamefont {Porter}}, \bibinfo {author} {\bibfnamefont
  {M.}~\bibnamefont {Rensink}},\ and\ \bibinfo {author} {\bibfnamefont
  {G.}~\bibnamefont {Smith}},\ }\href@noop {} {\bibfield  {journal} {\bibinfo
  {journal} {Contributions to plasma physics}\ }\textbf {\bibinfo {volume}
  {34}},\ \bibinfo {pages} {362} (\bibinfo {year} {1994})}\BibitemShut
  {NoStop}%
\bibitem [{\citenamefont {Simonini}\ \emph {et~al.}(1994)\citenamefont
  {Simonini}, \citenamefont {Corrigan}, \citenamefont {Radford}, \citenamefont
  {Spence},\ and\ \citenamefont {Taroni}}]{simonini1994}%
  \BibitemOpen
  \bibfield  {author} {\bibinfo {author} {\bibfnamefont {R.}~\bibnamefont
  {Simonini}}, \bibinfo {author} {\bibfnamefont {G.}~\bibnamefont {Corrigan}},
  \bibinfo {author} {\bibfnamefont {G.}~\bibnamefont {Radford}}, \bibinfo
  {author} {\bibfnamefont {J.}~\bibnamefont {Spence}},\ and\ \bibinfo {author}
  {\bibfnamefont {A.}~\bibnamefont {Taroni}},\ }\href@noop {} {\bibfield
  {journal} {\bibinfo  {journal} {Contributions to Plasma Physics}\ }\textbf
  {\bibinfo {volume} {34}},\ \bibinfo {pages} {368} (\bibinfo {year}
  {1994})}\BibitemShut {NoStop}%
\bibitem [{\citenamefont {Radford}\ \emph {et~al.}(1996)\citenamefont
  {Radford}, \citenamefont {Chankin}, \citenamefont {Corrigan}, \citenamefont
  {Simonini}, \citenamefont {Spence},\ and\ \citenamefont
  {Taroni}}]{radford1996}%
  \BibitemOpen
  \bibfield  {author} {\bibinfo {author} {\bibfnamefont {G.}~\bibnamefont
  {Radford}}, \bibinfo {author} {\bibfnamefont {A.}~\bibnamefont {Chankin}},
  \bibinfo {author} {\bibfnamefont {G.}~\bibnamefont {Corrigan}}, \bibinfo
  {author} {\bibfnamefont {R.}~\bibnamefont {Simonini}}, \bibinfo {author}
  {\bibfnamefont {J.}~\bibnamefont {Spence}},\ and\ \bibinfo {author}
  {\bibfnamefont {A.}~\bibnamefont {Taroni}},\ }\href@noop {} {\bibfield
  {journal} {\bibinfo  {journal} {Contributions to Plasma Physics}\ }\textbf
  {\bibinfo {volume} {36}},\ \bibinfo {pages} {187} (\bibinfo {year}
  {1996})}\BibitemShut {NoStop}%
\bibitem [{\citenamefont {Braams}(1996)}]{braams1996}%
  \BibitemOpen
  \bibfield  {author} {\bibinfo {author} {\bibfnamefont {B.}~\bibnamefont
  {Braams}},\ }\href@noop {} {\bibfield  {journal} {\bibinfo  {journal}
  {Contributions to Plasma Physics}\ }\textbf {\bibinfo {volume} {36}},\
  \bibinfo {pages} {276} (\bibinfo {year} {1996})}\BibitemShut {NoStop}%
\bibitem [{\citenamefont {Wiesen}\ \emph {et~al.}(2015)\citenamefont {Wiesen},
  \citenamefont {Reiter}, \citenamefont {Kotov}, \citenamefont {Baelmans},
  \citenamefont {Dekeyser}, \citenamefont {Kukushkin}, \citenamefont {Lisgo},
  \citenamefont {Pitts}, \citenamefont {Rozhansky}, \citenamefont {Saibene}
  \emph {et~al.}}]{wiesen2015}%
  \BibitemOpen
  \bibfield  {author} {\bibinfo {author} {\bibfnamefont {S.}~\bibnamefont
  {Wiesen}}, \bibinfo {author} {\bibfnamefont {D.}~\bibnamefont {Reiter}},
  \bibinfo {author} {\bibfnamefont {V.}~\bibnamefont {Kotov}}, \bibinfo
  {author} {\bibfnamefont {M.}~\bibnamefont {Baelmans}}, \bibinfo {author}
  {\bibfnamefont {W.}~\bibnamefont {Dekeyser}}, \bibinfo {author}
  {\bibfnamefont {A.}~\bibnamefont {Kukushkin}}, \bibinfo {author}
  {\bibfnamefont {S.}~\bibnamefont {Lisgo}}, \bibinfo {author} {\bibfnamefont
  {R.}~\bibnamefont {Pitts}}, \bibinfo {author} {\bibfnamefont
  {V.}~\bibnamefont {Rozhansky}}, \bibinfo {author} {\bibfnamefont
  {G.}~\bibnamefont {Saibene}}, \emph {et~al.},\ }\href@noop {} {\bibfield
  {journal} {\bibinfo  {journal} {Journal of nuclear materials}\ }\textbf
  {\bibinfo {volume} {463}},\ \bibinfo {pages} {480} (\bibinfo {year}
  {2015})}\BibitemShut {NoStop}%
\bibitem [{\citenamefont {Rognlien}\ \emph {et~al.}(2018)\citenamefont
  {Rognlien}, \citenamefont {Rensink},\ and\ \citenamefont
  {Stotler}}]{rognlien2018}%
  \BibitemOpen
  \bibfield  {author} {\bibinfo {author} {\bibfnamefont {T.}~\bibnamefont
  {Rognlien}}, \bibinfo {author} {\bibfnamefont {M.}~\bibnamefont {Rensink}},\
  and\ \bibinfo {author} {\bibfnamefont {D.}~\bibnamefont {Stotler}},\
  }\href@noop {} {\bibfield  {journal} {\bibinfo  {journal} {Fusion Engineering
  and Design}\ }\textbf {\bibinfo {volume} {135}},\ \bibinfo {pages} {380}
  (\bibinfo {year} {2018})}\BibitemShut {NoStop}%
\bibitem [{\citenamefont {Rambo}\ and\ \citenamefont
  {Denavit}(1994)}]{rambo1994}%
  \BibitemOpen
  \bibfield  {author} {\bibinfo {author} {\bibfnamefont {P.}~\bibnamefont
  {Rambo}}\ and\ \bibinfo {author} {\bibfnamefont {J.}~\bibnamefont
  {Denavit}},\ }\href@noop {} {\bibfield  {journal} {\bibinfo  {journal}
  {Physics of Plasmas}\ }\textbf {\bibinfo {volume} {1}},\ \bibinfo {pages}
  {4050} (\bibinfo {year} {1994})}\BibitemShut {NoStop}%
\bibitem [{\citenamefont {Rambo}\ and\ \citenamefont
  {Procassini}(1995)}]{rambo1995}%
  \BibitemOpen
  \bibfield  {author} {\bibinfo {author} {\bibfnamefont {P.}~\bibnamefont
  {Rambo}}\ and\ \bibinfo {author} {\bibfnamefont {R.}~\bibnamefont
  {Procassini}},\ }\href@noop {} {\bibfield  {journal} {\bibinfo  {journal}
  {Physics of Plasmas}\ }\textbf {\bibinfo {volume} {2}},\ \bibinfo {pages}
  {3130} (\bibinfo {year} {1995})}\BibitemShut {NoStop}%
\bibitem [{\citenamefont {Ghosh}\ \emph {et~al.}(2019)\citenamefont {Ghosh},
  \citenamefont {Chapman}, \citenamefont {Berger}, \citenamefont {Dimits},\
  and\ \citenamefont {Banks}}]{ghosh2019}%
  \BibitemOpen
  \bibfield  {author} {\bibinfo {author} {\bibfnamefont {D.}~\bibnamefont
  {Ghosh}}, \bibinfo {author} {\bibfnamefont {T.~D.}\ \bibnamefont {Chapman}},
  \bibinfo {author} {\bibfnamefont {R.~L.}\ \bibnamefont {Berger}}, \bibinfo
  {author} {\bibfnamefont {A.}~\bibnamefont {Dimits}},\ and\ \bibinfo {author}
  {\bibfnamefont {J.}~\bibnamefont {Banks}},\ }\href@noop {} {\bibfield
  {journal} {\bibinfo  {journal} {Computers \& Fluids}\ }\textbf {\bibinfo
  {volume} {186}},\ \bibinfo {pages} {38} (\bibinfo {year} {2019})}\BibitemShut
  {NoStop}%
\bibitem [{\citenamefont {Kolmes}\ \emph
  {et~al.}(2021{\natexlab{b}})\citenamefont {Kolmes}, \citenamefont {Ochs},\
  and\ \citenamefont {Fisch}}]{MITNS}%
  \BibitemOpen
  \bibfield  {author} {\bibinfo {author} {\bibfnamefont {E.}~\bibnamefont
  {Kolmes}}, \bibinfo {author} {\bibfnamefont {I.}~\bibnamefont {Ochs}},\ and\
  \bibinfo {author} {\bibfnamefont {N.}~\bibnamefont {Fisch}},\ }\href
  {https://doi.org/https://doi.org/10.1016/j.cpc.2020.107511} {\bibfield
  {journal} {\bibinfo  {journal} {Computer Physics Communications}\ }\textbf
  {\bibinfo {volume} {258}},\ \bibinfo {pages} {107511} (\bibinfo {year}
  {2021}{\natexlab{b}})}\BibitemShut {NoStop}%
\bibitem [{\citenamefont {{van Leer}}(1979)}]{VANLEER1979101}%
  \BibitemOpen
  \bibfield  {author} {\bibinfo {author} {\bibfnamefont {B.}~\bibnamefont {{van
  Leer}}},\ }\href
  {https://doi.org/https://doi.org/10.1016/0021-9991(79)90145-1} {\bibfield
  {journal} {\bibinfo  {journal} {Journal of Computational Physics}\ }\textbf
  {\bibinfo {volume} {32}},\ \bibinfo {pages} {101} (\bibinfo {year}
  {1979})}\BibitemShut {NoStop}%
\end{thebibliography}
\providecommand{\noopsort}[1]{}\providecommand{\singleletter}[1]{#1}%

\end{document}